\documentclass{article}

\usepackage[utf8]{inputenc}
\usepackage{multirow} 
\usepackage{rotating} 
\usepackage{bigstrut} 

\usepackage{geometry}
\usepackage{bbm}
\usepackage{upgreek}
\usepackage{setspace}
\usepackage{graphicx}
\usepackage{amssymb}
\usepackage{subcaption}
\usepackage{tikz}
\usepackage[font=footnotesize,labelfont=bf]{caption}
\usepackage{amsmath}
\usepackage{systeme}
\usepackage{multicol}
\usepackage{hyperref}\hypersetup{colorlinks = true, citecolor = blue, linkcolor=blue}

\newgeometry{left=1.86in,right=1.86in,top=1.73in,bottom=0.6in}

\begin{document}
\title{\vspace{-4.85cm}\textbf{\hspace*{-2.08cm}{\mbox{{\Large Magneto-optical-trap loading in a large optical-access experiment}}}}}
\author{
\vspace{0cm}\hspace*{-2.45cm}{\normalsize M. Gaudesius$^{1,2,\footnote{\hspace{0cm}\mbox{Present address: The Pennsylvania State University, University Park, Pennsylvania 16802, USA} \\\hspace*{11.2pt}$^\dagger$Contact author: gd.mar900@gmail.com\\\hspace*{11.2pt}$^\ddagger$Contact author: biedermann@ou.edu}\;,^\dagger}$ , J. M. Lee$^{1,2}$, L. A. Kraft$^{1,2}$, J. C. Gordon$^{1,2}$, and G. W. Biedermann$^{1,2,^\ddagger}$} \\
\hspace*{-2.94cm}\vspace{-0.1cm}{\small\textit{ $^{1}$Homer L. Dodge Department of Physics and Astronomy, The University of Oklahoma, Norman, Oklahoma 73019, USA }} \\ 
\hspace*{-2.72cm}\vspace{-0.1cm}{\small\textit{ $^{2}$Center for Quantum Research and Technology, The University of Oklahoma, Norman, Oklahoma 73019, USA }} \\
 }
\date{}
\maketitle

\vspace{-0.45cm}
\leftskip=-2.08cm\rightskip=-2.08cm
{\small We present an experimental, numerical, and analytical study of strontium magneto-optical trap (MOT) loading from a cold atomic beam in a configuration optimized for high numerical aperture optical tweezers. Our approach orients the beam flow along the MOT symmetry axis to reduce the experimental complexity and maximize the overall optical access into the scientific region of study. We use a moving molasses technique to enable this configuration and show that its performance depends critically on metastable-state shelving (to $5s5p\,^3\!P_2$) during the atom transfer to the three-dimensional (3D) MOT. Furthermore, we find that the parameters for optimal transfer efficiency are bounded by dark-state loss (to $5s5p\,^3\!P_0$) in the trap region where repumping is present. These observations are verified to great degree of accuracy using both our developed analytical and numerical models. The corresponding 3D simulation tool is used to perform a comprehensive study of the trap loading dynamics, beginning at the oven exit and ending at the 3D MOT, demonstrating its effectiveness in optimizing an effusive oven experiment. 
}
\vspace{10pt}
\begin{multicols}{2}\setlength{\columnsep}{2pt}

\leftskip=-0.5cm
\section{\hspace{-0.38cm}Introduction}\label{sec:Introduction}
\leftskip=-3cm\rightskip=0.15cm
\vspace{-7pt}
Both alkali and alkaline-earth metal atoms form an important part of cold atom experiments owing to easily accessible transition frequencies and effectively closed cooling cycles. The latter atoms, possessing two valance electrons instead of one, have unique energy level structures that have attracted a growing interest in the atomic physics community. For example, a bosonic isotope of strontium, $^{88}$Sr, offers a pure ground state and multiple narrow transitions, making it an outstanding candidate in several areas of research, such as atom interferometry \cite{1,2}, optical frequency standards \cite{3,4}, tests of fundamental physics \cite{5,6}, as well as quantum simulation and computation with cold Rydberg atoms \cite{7,8,9}. These kinds of investigations generally are based on techniques that are widely used and established. Particularly for the Rydberg-atom based simulation and computing, the manipulation and control of atoms relies heavily on the usage of optical tweezers \cite{10}-\cite{13}. In such experiments, it is often desirable to achieve a sizable atomic array \cite{14,15}, a high detection efficiency via large solid-angle light collection \cite{16,17}, and an extensive trap lifetime \cite{18,19}.

Modern experiments commonly accomplish this by using high numerical aperture (NA) microscope objectives to create and detect a tweezer array, with the atoms loaded from a magneto-optical trap (MOT) in UHV conditions. The high NA objectives consume precious optical access, often leading to a departure from the optimal MOT laser beam configuration \cite{20,21}. Loading a MOT from a cold source achieves UHV conditions and increased experimental repetition rates \cite{22,23,24}, but further consumes lines of access into the scientific study region of the apparatus. We propose and demonstrate in this paper a technique for the MOT loading that solves this challenge in an effusive oven experiment. Hence, high NA optical tweezers can be realized while simultaneously opening additional laser pathways for atom array manipulation. We

\leftskip=0.15cm\rightskip=-3cm 
\noindent verify and optimize this approach to great accuracy with a simulation tool incorporating the physics deemed by our analytical theory to be critical for our main observations. 

\vspace*{4pt}
\hspace*{8pt}\includegraphics[scale=0.35]{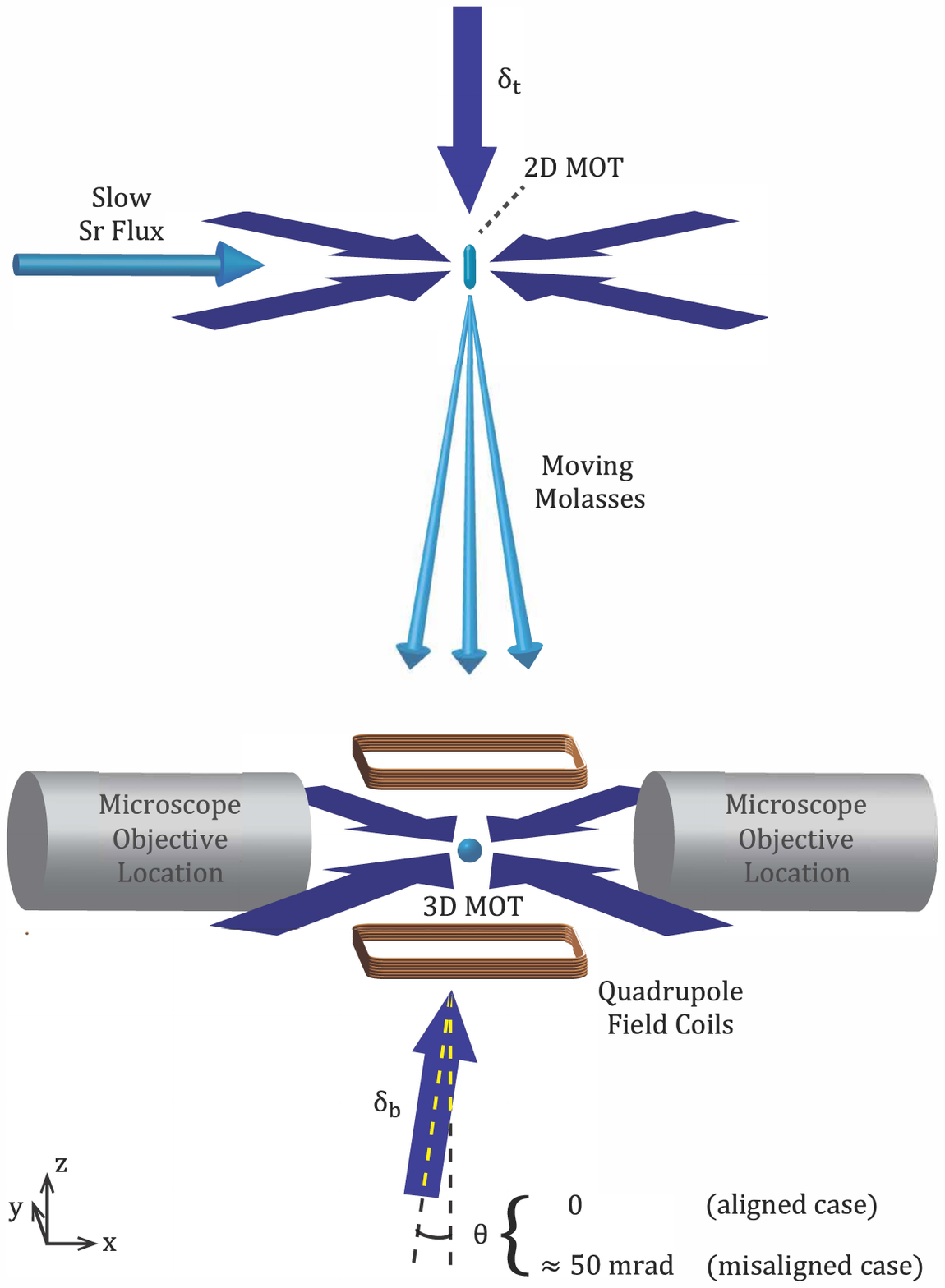}\label{fig:1}
\vspace*{-3pt}
\begin{footnotesize}
\begin{spacing}{1}
\vspace*{-3pt}\noindent{\textbf{Figure 1:} Moving molasses technique for transferring atoms between magneto-optical traps (MOTs) and enabling large optical access experiments (with proposed locations for microscope objectives). The 3D MOT loading relies on its axial beams creating an atomic molasses beam from the location of the 2D MOT (itself loaded by, e.g., a Zeeman-slowed Sr flux). We present two cases: (i) Aligned, where the top and bottom axial beams are counter propagating and must be unequally detuned ($\delta_\text{t}\neq\delta_\text{b}$); and (ii) Misaligned, where the bottom beam is slightly angled (to 3D MOT) for enhanced loading with $\delta_\text{t}\neq\delta_\text{b}$, in addition to opening the possibility to load given $\delta_\text{t}=\delta_\text{b}$. Note that the results of this work consider case (i) unless specified. \newline}
\end{spacing}
\end{footnotesize}

\leftskip=-3cm\rightskip=0.15cm
\normalsize
\vspace*{-15pt}
A variety of methods have been demonstrated for the MOT loading \cite{25}-\cite{29}. For high vapor pressure atoms such as rubidium or cesium, a MOT can be loaded directly from background vapor \cite{30,31}. On the other hand, for low vapor pressure atoms such as strontium or ytterbium, an effusive oven \cite{32,33} or other methods \cite{34:extra,34} must be used to achieve sufficient density. The effusive oven approach typically relies on collimating the atoms into a beam that is directed at a MOT. For reduced collisions, a two-stage MOT loading method can be taken, where the atoms from the original MOT (commonly, 2D) are transferred to the main MOT (3D) with the help of different techniques (see below). The transfer can be done through a differential pumping stage, leading to an improved vacuum in the science region \cite{35,36} and hence achieving a long trap lifetime required for the extended manipulation and detection of large atom arrays \cite{14,37,38}.

The effusive oven approach is not without its challenges. Heating the reservoir to a large temperature exceeding, e.g., $400\;^{\circ}$C for strontium requires regular refurbishments of the chamber components as well as heightened maintenance of the vacuum \cite{36,38,39}. Moreover, a Zeeman slower \cite{40,41} is required to collect a large fraction of the hot atoms to within the capture speed of a downstream MOT. In a two-stage MOT method, the loading efficiency of a 3D MOT is bounded by that of a 2D MOT and is critically dependent on the transfer technique that is employed.

A simple way of performing the transfer is by adding to the 2D MOT a single push beam that creates an atomic flow towards the 3D MOT \cite{35,36,42,43}. For enhanced flux, a 2D$^+$ MOT can be used, where a pair of Doppler-cooling ``+" beams are added orthogonally to the 2D MOT beams, and an atomic beam is created within a shadow of the ``+" beam that opposes the pushing ``+" beam \cite{22}. This shadow or extraction column is produced by a small aperture in the in-vacuum mirror that reflects light to create the opposing beam. This approach is similar to the low-velocity intense source (LVIS) system \cite{44}, where the atomic beam emerges from an aperture in a MOT mirror. The transfer technique employing a 2D$^+$ MOT achieves the flow with the ``+" beams having different intensities, while the LVIS system uses the same beam parameters but results in weaker extraction. In contrast to these approaches, we utilize differently detuned 3D MOT axial beams and no in-vacuum assistance (see Fig. \hyperref[fig:1]{1}), allowing for relatively straightforward assembly of our system and modeling of the transfer from our 2D MOT. We note that it is not called a 2D$^+$ MOT to stress that the axial beams are part of the 3D MOT system and that the axial beam misalignment is a viable option for transfer (misaligned case in Fig. \hyperref[fig:1]{1}). In the past \cite{45}, the 

\leftskip=0.15cm\rightskip=-3cm 
\noindent misalignment of 2D MOT beams themselves has been used to achieve the transfer into the 3D MOT region but without improved optical access to this main region like in the case of all the previous techniques. Overall, our approach creates broad optical access to the 3D MOT that uses convenient beam geometry achieved by having its axial beams creating the flow from the 2D MOT while excluding in-vacuum parts. 

The optimization of our approach requires detailed considerations of an asymmetric MOT configuration with complex loading dynamics involving metastable-state shelving and dark-state loss. Hence, a pivotal goal of this paper is to provide a comprehensive simulation tool addressing the dynamics associated with the loading of MOTs, beginning at the hot source and ending at the 3D MOT. An analytical theory that we develop provides a clear-cut physical picture behind our 3D MOT loading approach.

We start in Sec. \hyperref[sec:Experiment]{II} by giving an overview of our effusive strontium-oven experimental setup that is optimized for high NA optical tweezers and discuss the corresponding numerical setup. It is based on a model accounting for light polarization, arbitrary beam orientations in 3D space, and atom lifetime. In Sec. \hyperref[sec:MainResults]{III}, we describe our technique for populating the main science region and present the main measurement results with comparison to our numerical and analytical models. In Sec. \hyperref[sec:Conclusions]{IV}, we conclude and discuss the future prospects of our work. Appendixes \hyperref[sec:Appendix_A1]{A} and \hyperref[sec:Appendix_A2]{B} cover theoretical aspects of our numerical model, and Appendix \hyperref[sec:Appendix_B]{C} provides analytical calculations in support of our main observations. Additionally, the Supplemental Material \cite{2:0} includes a numerical study of the loading, with general guidelines for optimizing an effusive oven experiment.

\leftskip=0.15cm\rightskip=-3cm
\leftskip=0.62cm
\section{\hspace*{-0.38cm}\mbox{Experimental and numerical}\\\hspace*{2.0cm}\mbox{setups}}\label{sec:Experiment}
\leftskip=0.15cm\rightskip=-3cm

Our experiment employs the bosonic isotope $^{88}$Sr (mass $m=1.46\cdot10^{-25}$ kg), operating on the blue transition $5s^2\,^1\!S_0\rightarrow5s5p\,^1\!P_1$ with the wavelength $\lambda=461$ nm, the linewidth $\Gamma=2\pi\times32$ MHz, and the saturation intensity $I_{sat}=42.72$ mW/cm$^2$. The effusive atomic beam cooling, and the magneto-optical trapping together with the moving molasses transfer are thus realized with radiation pressure from this $^1\!S_0\;{\rightarrow}\;^1\!P_1$ transition, as discussed in detail below. The atoms exposed to the blue light can shelve to the metastable stable $5s5p\,^3\!P_2$ (reached via the decay channel $5s5p\,^1\!P_1\rightarrow5s4d\,^1\!D_2\rightarrow\,5s5p$ $^3\!P_2$), which we find to be critical for achieving an efficient transfer into the final trapping location (from the\;\,2D\;\;MOT\;\,into

\leftskip=-3cm\rightskip=0.15cm
\begin{figure*}
\vspace*{-112pt}
\hspace*{-89pt}\includegraphics[scale=0.54]{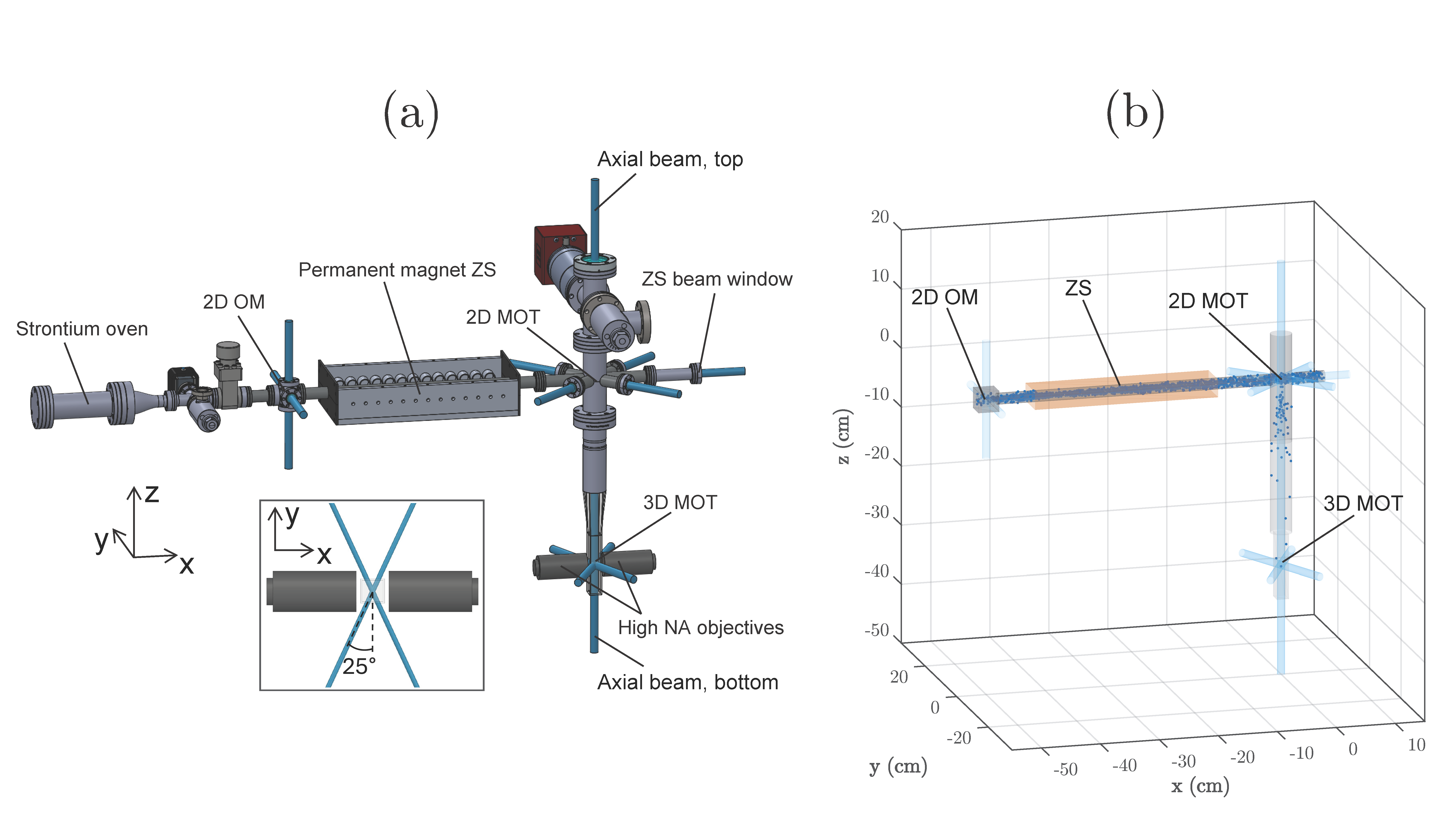}
\vspace*{-22pt}
 \captionsetup{width=1.49\linewidth}
  \caption*{\textbf{Figure 2:} (a) Schematic view of the experimental vacuum chamber (OM: optical molasses; ZS: Zeeman slower; MOT: magneto-optical trap). The inset displays a top view of a high numerical aperture (NA) tweezer setup; the 3D MOT side beams are mirrored with respect to the radial axis perpendicular to the objectives axis at a $25\,^{\circ}$ angle. Note that the permanent magnets and the coils producing, respectively, the 2D and 3D MOT magnetic fields are not shown. (b) The corresponding numerical setup, including the 2D OM, ZS, 2D and 3D MOTs. The blue dots are superparticles. A video version of this figure is available as online Supplemental Material \cite{2:0}.}
\label{fig:2}
\end{figure*}

\leftskip=-3cm\rightskip=0.15cm
\noindent the 3D MOT). Repumping is employed only at the 3D MOT and tuned to the cyan transition $5s5p\,^3\!P_2\rightarrow5p^2\,^3\!P_2$ with the wavelength $\lambda_c=481$ nm, the linewidth $\Gamma_c=2\pi\times14$ MHz, and the saturation intensity $I_{sat,c}=16.41$ mW/cm$^2$. This leads to the shelved atom revival, increasing the 3D MOT lifetime \cite{35,2:1,2:2} and enhancing the detection fidelity in optical tweezer experiments \cite{21,2:3}. Note that a loss into the long-lived dark state $5s5p\,^3\!P_0$ can occur during the repumping, bounding the lifetime increase \cite{2:1,2:2}. 

The experimental setup is seen in Fig.  \hyperref[fig:2]{2}(a). Its main stages include an effusive oven, a 2D optical molasses (OM), a Zeeman slower (ZS), a 2D MOT, and, lastly, a 3D MOT. The atoms originating from the effusive oven are first collimated to form a beam that is cooled transversely by the 2D OM and longitudinally by the ZS. The longitudinal cooling slows the atoms into a velocity class within the capture range of the 2D MOT, with the transverse cooling providing additional collimation for improved loading efficiency. The 3D MOT is loaded from the 2D MOT by using the axial beams only (discussed in Sec. \hyperref[sec:MainResults]{III}), granting the optical access for the high NA tweezer objectives. We use a 240 l/s ion-getter pump (SAES NEXTorr Z 200) that we attach above the 2D MOT chamber. Without the getter activated, it maintains an estimated $10^{-9}$ Torr pressure at the main science region. With this pressure, the mean lifetime due to background collisions is expected to be on the order on 1 s \cite{2:4}, which we deem satisfactory for our investigations. We determine the atom numbers at various chamber locations (2D OM, 2D MOT, and 3D MOT) from the voltage readings of an avalanche photodiode (Hamamatsu C12703-01) 

\leftskip=0.15cm\rightskip=-3cm
\noindent together with calculated scattering rates and detection solid angles \cite{32,2:6}. 

The numerical setup is seen in Fig. \hyperref[fig:2]{2}(b). It mimics the experimental one starting from the 2D OM stage, with its initial numerical conditions dictated by the collimation geometry after the oven exit and the measured flow rate (Sec. \hyperref[sec:Experiment_i]{II.A}). We base our model on the $F=0\rightarrow F'=1$ transition, faithful to the blue trapping descriptions of $^{88}$Sr and allowing for a proper treatment of the features related to the magnetic field and light polarization. We build upon the model in Ref. \cite{2:7} by considering arbitrary beam orientations, although we currently exclude its multiple-scattering effects. The theoretical aspects are detailed in Appendixes \hyperref[sec:Appendix_A1]{A} and \hyperref[sec:Appendix_A2]{B}, where the radiation pressure and the atom lifetime are respectively treated.

The dynamics are numerically implemented using the Leapfrog algorithm that updates the velocities and positions of the superparticles (collections of regular particles) once the forces are computed (scaled by the number of regular particles represented by one superparticle) \cite{2:7,2:7_extra}. To conserve computational resources, the simulation is divided into three parts (with the first two determining the initial conditions for their respective next part): (i) 2D OM, (ii) ZS and 2D MOT, and (iii) 2D and 3D MOTs. [Figure \hyperref[fig:2]{2}(b) showcases the simulation after (i), with (ii) and (iii) performed simultaneously.] We use up to $10^7$ superparticles, in that way obtaining convergent results, and release them continuously according to the given loading rates (at the simulation end, all the superparticles have been released). Each superparticle has a lifetime calculated by multiplying the mean lifetime (Eq. \hyperref[eq:tau_simple]{B1}) with a number predrawn from a unit exponential dis-
\end{multicols}

\newgeometry{left=1.86in,right=1.86in,top=0.85in,bottom=0.6in}
\begin{multicols}{2}\setlength{\columnsep}{2pt}
\leftskip=-3cm\rightskip=0.15cm
\noindent tribution. Once the lifetime is over in a region without the repumping (the enhancement factor $\mathcal{E}=1$  in Eq. \hyperref[eq:tau_simple]{B1}), a superparticle is affected only by gravity ($-\textsf{z}$ direction), in accordance with it being shelved to the metastable state $5s5p\,^3\!P_2$. Given a superparticle drifts into the repumping region, it is revived with a new lifetime (with $\mathcal{E}=27$) after which it is deleted from the simulation as it is assumed to be lost to the dark state $5s5p\,^3\!P_0$. Note that this new lifetime assumes an enhancement ($\mathcal{E}$) that is approximately maximal for our employed repumping scheme and is independent of the atomic density (which changes during MOT loading) \cite{2:2}. Note also that the superparticles are deleted at the chamber bounds in accordance with the surface binding energy of strontium being especially large \cite{2:8}, and the beams naturally have a finite diameter as set by the size of the optics apertures (half-inch). In the below subsections, we discuss in detail the experimental and numerical setups. 

\leftskip=-2.5cm
\subsection*{\hspace{1.61cm}\mbox{A. Strontium oven and}\\ \hspace*{2.0cm}\mbox{2D optical molasses}}\label{sec:Experiment_i}
\leftskip=-3cm\rightskip=0.15cm

\vspace*{-2pt}
We operate the strontium oven at a relatively modest temperature of $440\,^{\circ}$C, in an effort to minimize the chamber degradation over time. We stack microcapillaries (at $500\,^{\circ}$C, higher to prevent clogging) collimating the atoms into a beam \cite{2:9} and use a differential tube (before the 2D OM) providing additional collimation. Each microcapillary is $l_{cap}=8$ mm long with the inner diameter of $d_{cap}=0.17$ mm, while the differential tube is $l_{dif}=44$ mm long with the inner diameter of $d_{dif}=6$ mm; the separation distance between the microcapillary stack and the differential tube is $\Delta l=204$ mm. These dimensions of our dual stage collimation affect the transverse velocity profile as discussed below. The flow rate measured at the location of the 2D OM is $2\times10^{11}$ atoms/s. The 2D OM itself is realized by two retroreflected beams with an individual power of 6 mW, a $1/e^2$ radius of $3.2$ mm, and a detuning of $-30$ MHz. 

The initial conditions for the 2D OM simulation are set by (i) the oven temperature; (ii) the atomic flow rate; and (iii) the collimation divergence angle calculated using the dual stage collimation length $l_{tot}\equiv l_{cap}+l_{dif} + \Delta l$ and the end opening size $d_{dif}$, $\theta_{d}=\text{tan}\left( \frac{d_{dif}}{l_{tot}} \right) \approx 23$ mrad (which by construction lies between the maximum divergence angles from the capillary start and end, respectively, $\text{tan}\left( \frac{d_{cap}}{l_{cap}} \right) \approx 21$ mrad and $\text{tan}\left( \frac{d_{dif}}{l_{tot}-l_{cap}} \right) \approx 24$ mrad). The flow rate is our only \textit{free} numerical parameter, because it is entirely dependent on a measurement. The divergence-angle distribution for the superparticles is a combination of (i) a normal distribution with $\mu_{d}=2\sigma_{d}=\theta_{d}/2$ ($\mu_{d}$ and $\sigma_{d}$ are the mean and standard deviation) and cut off outside 0 and $\theta_{d}$, and (ii) a uniform distribution that lies between these cut off values and that is relevant only for the superparticles where the initial normal distribution draw exceeds these cut offs. A given divergence angle from the distribution finally\;\,deter-

\leftskip=0.15cm\rightskip=-3cm
\noindent mines a superparticle's initial transverse velocity from the total velocity (sum of longitudinal and transverse) that is drawn from the Maxwell-Boltzmann distribution evaluated at the oven temperature. The effect of the 2D OM is evaluated at the 2D MOT, as discussed in Sec. \hyperref[sec:Experiment_iii]{II.C}, while the numerical optimization is discussed in the Supplemental Material \cite{2:0}.

\leftskip=0.15cm\rightskip=-3cm
\vspace*{-153pt}
\hspace*{-80pt}\includegraphics[scale=0.64]{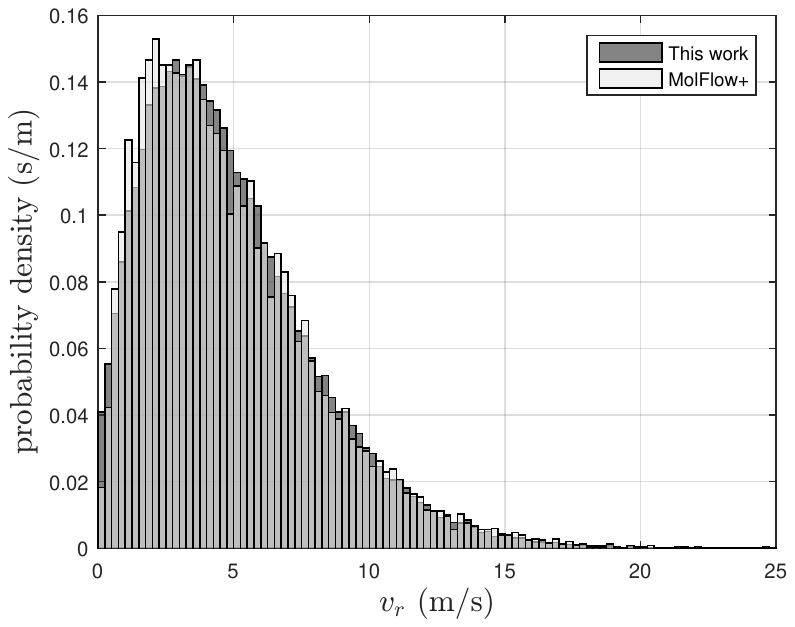}\label{fig:3}
\vspace*{-155pt}
\begin{footnotesize}
\begin{spacing}{1}
\vspace*{-3pt}\noindent{\textbf{Figure 3:} Numerically calculated radial velocity $v_r$ distributions after our dual stage collimation for strontium particles emerging from an oven. The dark gray histogram is obtained using our divergence-angle assumption, whereas the light gray histogram is obtained with MolFlow+ \cite{2:MolFlow+}. See the main text for details. \newline}
\end{spacing}
\end{footnotesize}

To justify the choice of our divergence-angle distribution, we use MolFlow+ \cite{2:MolFlow+} to perform a simulation of oven-particle trajectories. This simulation is initiated inside of a cube (20 mm edge length) representing an oven, and the trajectories are let to evolve through our dual stage collimation geometry until a total of $10^4$ events are recorded at the output. The respective temperatures of the oven and the capillaries are as stated above ($440\,^{\circ}$C and $500\,^{\circ}$C), while the differential tube is at room temperature (as in the experiment). The initial velocities are drawn from the Maxwell-Boltzmann distribution, and the angles of reflection by surfaces are drawn from the cosine distribution (according to Knudsen's cosine law). No sticking to surfaces is assumed, but hitting the tube results in practically infinite dwell-time due to the much smaller tube temperature. This effectively filters away the trajectories with divergence angles larger than those resulting from reflections right off the periphery of a given capilary and continuing near the tube end edge. We note that the corresponding greatest angle is roughly equal to our own model's upper cutoff (comparing ${\approx}24$ to ${\approx}23$ mrad). In Fig. \hyperref[fig:3]{3}, we compare the distributions for the output radial velocities $v_r$ obtained using our divergence-angle assumption (dark gray) and MolFlow+ (light gray). We observe that the two distributions are nearly identical (up to randomization error), proving that our simple assumption can accurately predict the physical velocities for our geometry output. 

\end{multicols}

\newgeometry{left=1.86in,right=1.86in,top=1.73in,bottom=0.6in}
\begin{multicols}{2}\setlength{\columnsep}{2pt}

\leftskip=-3cm\rightskip=0.15cm
\leftskip=-0.4cm
\subsection*{\hspace{0.1cm}B. Zeeman slower}\label{sec:Experiment_ii}
\leftskip=-3cm\rightskip=0.15cm

The ZS beam enters through a heated sapphire window (at 200$\,^{\circ}$C, to prevent deposition of strontium), as seen at the right of Fig. \hyperref[fig:2]{2}(a). It has a power of 70 mW (taking into account a non-negligible loss through the window), a $1/e^2$ radius of $4$ mm, and a detuning of $-370$ MHz. The slower design is based on using permanent magnets that are placed inside of cups surrounding opposite sides of the nipple connecting the 2D OM and 2D MOT chambers (50 cm apart) \cite{41}. It produces a transverse magnetic field ($\textsf{xy}$ plane) that increases non-monotonically from $-400$ G to $0$ G over the total length of 27 cm. We note that an iron plate attached at the slower exit provides a magnetic shielding of the sensitive end field (mainly against the 2D MOT field).

For our simulations, we have derived a heuristic model (Eq. \ref{eq:B_Zeeman}), where the ZS magnetic field is pieced together by linear segments with given start and end point values. Note that due to its transverse nature, the inclusion of light polarization is essential for a correct description of the slowing. For a smooth profile that closely matches the experimental one, we choose $n_S=30$ segments in our simulations. In the Supplemental Material \cite{2:0}, we outline a method for the numerical optimization, which we note has not been implemented in the present case. Nonetheless, we expect it to yield accurate optimization results, given that the heuristic model for the ZS magnetic field leads to comparable numerical and experimental 2D MOT loading rates discussed below.

\subsection*{\hspace{1.5cm}C. 2D magneto-optical trap}\label{sec:Experiment_iii}

To realize the 2D MOT, we use permanent magnets creating a $67$ G/cm field gradient along the axes of two retroreflected beams with an individual power of 20 mW, a $1/e^2$ radius of $3.2$ mm, and a detuning of $-32$ MHz. The magnets are placed inside of holders [not shown in Fig. \hyperref[fig:2]{2}(a)] above and below ($\textsf{z}$ axis) the atomic beam entrance and exit nipples of the 2D MOT chamber, with the magnetization directions being opposite ($\textsf{y}$ axis) on the respective nipples. The resulting magnetic field is zero along the axis perpendicular to the main 2D MOT plane ($\textsf{xy}$ plane) as well as non-cylindrically-symmetric around this axis (unlike in a 3D MOT). The atoms are transferred into the 3D MOT along this axis (Sec. \hyperref[sec:Experiment_iv]{II.D}). 

The permanent magnet field of the 2D MOT is modeled accordingly by Eq. \ref{eq:B_2DMOT}, where we include the experimental distances between the magnets (modeled as dipoles), their volumes ($V=1''\times1/2''\times1/4''$ each) and remanent magnetization ($B_0=1.44$ T; N52 grade neodymium magnet). Due to\;\;the\;\;particular\;\;asymmetry

\leftskip=0.15cm\rightskip=-3cm
 \noindent of the field in the main 2D MOT plane, the inclusion of light polarization is essential for a correct description of the 2D MOT loading. We report in Tab. \hyperref[tab:1]{1} the loading rates obtained experimentally and numerically to evaluate the effects of the 2D OM and the ZS magnets.

\vspace*{8pt}\hspace*{-2pt}\begin{tabular}{ |c|c|c|c| } 
\hline
\multicolumn{4}{|c|}{2D MOT loading rates (atoms/s $\times10^8$)} \\
 \hline
 Configuration & $\mathcal{F}_{exp}$ & $\mathcal{F}_{sim}$ & $1-\mathcal{F}_{sim}/\mathcal{F}_{exp}$ \\ 
 \hline
 $-$ ZS, $-$ 2D OM & $1.6$ & $1.3$ & $19\,\%$ \\ 
 \hline
 $-$ ZS, $+$ 2D OM & $3.6$ & $2.1$ & $42\,\%$ \\ 
 \hline
 $+$ ZS, $-$ 2D OM & $7.1$ & $9$ &  $-27\,\%$ \\ 
 \hline
 $+$ ZS, $+$ 2D OM & $15$ & $13$ & $13\,\%$ \\ 
 \hline
\end{tabular}\label{tab:1}
\begin{footnotesize}
\begin{spacing}{1}
\vspace*{7pt}\noindent{\textbf{Table 1:} A comparison of the 2D MOT loading rates obtained experimentally ($\mathcal{F}_{exp}$) and numerically ($\mathcal{F}_{sim}$). ``$+$" and ``$-$" refer to the 2D OM or the ZS magnets being used and not being used, respectively.\newline}
\end{spacing}
\end{footnotesize}

\vspace*{-3pt}
Starting with row/case 1, where the atom dynamics are not affected by the 2D OM or the ZS magnets, the discrepancy between the experiment and simulation is less than $20\,\%$. This is noteworthy considering there is only one free numerical parameter (the total atomic flux emitted by the oven). The simulation is highly influenced by the initial velocity profile, and by employing our small-angle approximation for the oven (Sec. \hyperref[sec:Experiment_i]{II.A}) proves to yield closely matching results. The atom lifetime is also an important factor to consider (see Appendix \hyperref[sec:Appendix_A2]{B}), given that the atoms can get shelved to a metastable state as they travel through the Zeeman nipple and hence elude the 2D MOT capture. 

Comparing case 1 to cases 2 and 3, the effects of the 2D OM and the ZS magnets are respectively evaluated. The 2D OM is less efficient numerically (${\lesssim}40\,\%$), while the opposite is true for the ZS (${\gtrsim}30\,\%$). For the 2D OM, the result can be indicative of a higher portion of atoms moving faster transversely in the experiment and thus being more susceptible to the transverse cooling. For the ZS, it can be indicative of an effect, such as the light attenuation, being of importance (omitted for simplicity); indeed, we expect a non-negligible ZS beam attenuation given the modest flow rate and the long ZS nipple, resulting in an optically thick atomic medium. The attenuation would impact the lifetime of the atoms, in addition to the dynamics based on the radiation pressure alone, thus affecting their passage along the ZS.

Case 4, which is employed in our main simulations (Sec. \hyperref[sec:MainResults]{III}), combines both the 2D OM and the ZS magnets. We obtain a near-quantitative agreement between the experiment and simulation, with the discrepancy being less than $15\,\%$. Apart from the effectiveness of the underlying assumptions of the model, this is partially due to the higher loading rate prediction with the ZS magnets compensating for the lower one with the 2D OM. 

\leftskip=-0.2cm
\subsection*{\hspace{-1.25cm}D. 3D magneto-optical trap}\label{sec:Experiment_iv}
\leftskip=-3cm\rightskip=0.15cm

The 3D MOT field is realized with a pair of rectangular anti-Helmholtz coils surrounding the glass cell at the main science region [coils indicated by Fig. \hyperref[fig:1]{1} but not shown in Fig. \hyperref[fig:2]{2}(a); $2\,\text{cm}\,\times\,2.5\,\text{cm}$ radial ($\textsf{x}\times\textsf{y}$) cell dimensions], resulting in an approximately cylindrically-symmetric field around the transfer axis ($\textsf{z}$ axis) at the MOT center ($31.5$ cm below the 2D MOT center). Along this axis, we have two independent beams, each with a power of 3.5 mW and a $1/e^2$ radius of $3.2$ mm; one stems from below the 3D MOT with a fixed detuning of $-46$ MHz and the other counter propagates from above the 2D MOT with a detuning that we vary for testing our loading technique (Sec. \hyperref[sec:MainResults]{III}). We note that we also consider a special configuration where the bottom beam is slightly misaligned, as discussed later. In the radial plane ($\textsf{xy}$ plane), there are two retroreflected beams with an individual power of 2 mW, a $1/e^2$ radius of $2.4$ mm, and a detuning of $-30$ MHz. They are lowered by approximately one beam radius with respect to the quadrupole field zero, resulting in increased loading efficiency (discussed in detail in Sec. \hyperref[sec:MainResults]{III}). Additionally, mirroring them at $25\,^{\circ}$ with respect to the radial axis perpendicular to the objectives axis (i.e., with respect to the $\textsf{y}$ axis) grants the necessary optical access for the high NA optical tweezers while simultaneously covering most of the science region for enhanced loading (see the Supplemental Material \cite{2:0} for the numerical verification). One of the retroreflected pairs also contains the repumping light with a total power of $8$ mW; we find the repumping to be critical as most of the atoms get shelved to the metastable state $5s5p\,^3\!P_2$ before reaching the science region. 

The 3D MOT magnetic field is modeled by Eq. \ref{eq:B_3DMOT}, which is an approximation to the field created by the rectangular coils used in our experiment. Note that the modeled field is present in the transfer region (the same is true for the 2D MOT field); this information feeds into the scattering cross sections (Eq. \ref{eq:sigma}), thus affecting the atom dynamics. The radiation pressure forces corresponding to the radial plane beams take into account the lowering (1 radial beam radius) as well as the angling ($25\,^{\circ}$). The repumping is present in one of the radial beam pairs, like in the experiment. As mentioned in the introduction of Sec. \hyperref[sec:Experiment]{II}, when a given superparticle drifts into the repumping region, it is revived with a new lifetime that is enhanced compared to the non repumping region. 

\leftskip=-0.6cm
\section{\hspace{-0.38cm}\mbox{Main results}}\label{sec:MainResults}
\leftskip=-3cm\rightskip=0.15cm

Our 3D MOT loading involves a simple technique depicted\;\;in\;\;Fig.\;\;\hyperref[fig:1]{1}:\;\;The\;\;axial\;\;3D MOT\;\;laser\;\;beams\;\,are

\leftskip=0.15cm\rightskip=-3cm
\noindent used to create an atomic molasses beam from the location of the 2D MOT, providing cold atoms for the loading. This atomic beam is the result of a radiation-pressure imbalance between the axial beams, which is attained by (i) using unequal detunings in case they are counter propagating or (ii) having the bottom beam slightly angled to the 3D MOT when equally detuned. In the latter case, introducing a detuning difference results in enhanced loading (discussed in Sec. \hyperref[sec:MainResults_iii]{III.C}). 

With the moving molasses technique, the 3D MOT can be loaded for a broad range of parameters (discussed below), as the atoms accumulate below the quadrupole field zero as set by the axial pressure imbalance. This technique creates an advantageous MOT laser beam configuration as compared to adding a push or cold flux beam, which may otherwise require operating the MOT beams at angles compromising the loading \cite{20,21}. Moreover, a broad optical access to the scientific study region is achieved, creating additional laser-pathways for atom array manipulation with a high NA optical tweezer setup.

Below in this section, we first develop an analytical model for the 3D MOT loading using our technique and then compare the main experimental observations to this model as well as the numerical model. Lastly, we discuss the configuration with a misaligned bottom beam. We note that this special configuration obfuscates analytical treatment and is thus not central to our present analysis.

\leftskip=0.6cm
\subsection*{\hspace{1.58cm}\mbox{A. The analytical model}}\label{sec:MainResults_i}
\leftskip=0.15cm\rightskip=-3cm

Here we develop an analytical theory for the 3D MOT loading, with Fig. \hyperref[fig:1]{1} providing a visual guide (the aligned case is treated). We perform a step-by-step analysis, starting with the 2D MOT region, then describe the molasses beam in the transfer region (between the 2D and 3D MOTs) and the atom capture in the 3D MOT, and finally obtain an equation for the atom number in the 3D MOT. The theory incorporates the atomic diffusion effect and the shelving to the metastable state $5s5p\,^3\!P_2$ during the transfer, as well as the loss to the dark state $5s5p\,^3\!P_0$ in the final trap region where repumping is present. Note that the lifetimes associated with these respective processes (shelving and loss) are intrinsically Poissonian (present in our numerical model), but here we choose average lifetimes for simplicity. Furthermore, we treat cut offs with Heaviside functions providing abrupt boundaries, which is a legitimate choice when considering dimensional constraints but crudely approximating transitions in a MOT that are naturally expected to be smooth. Achieving such smoothness through the complexification of our model is not necessary considering the close match with experiment (Sec. \hyperref[sec:MainResults_ii]{III.B}).

\end{multicols}

\newgeometry{left=1.86in,right=1.86in,top=1.65in,bottom=0.6in}
\begin{multicols}{2}\setlength{\columnsep}{2pt}

\leftskip=-3cm\rightskip=0.15cm 
Starting from the 2D MOT, the unequally detuned axial 3D MOT beams create an atomic molasses beam directed at the 3D MOT with the flow rate $\mathcal{F}_0$. In addition to having a strong longitudinal velocity component, the atoms in this beam naturally possess a transverse velocity distribution impacting their capture probability in the 3D MOT.

As the atoms continue downstream in the transfer region, they accumulate a radial position spread due to the diffusion stemming from the radiation pressure of the axial (transfer) beams. This spread can be affected by the shelving to the metastable state, which freezes the diffusive velocity distribution. To obtain the transfer probability $\mathcal{P}_\oplus$ into the 3D MOT due to the randomness in the atomic beam, we take the radial spread to be normally distributed with the variance $\sigma_r^2$ and integrate this distribution as follows. Two different regimes are considered with their corresponding integration limits, in order to accommodate for the substantial range of magnetic field gradients explored in the experiment. In the first regime, we assume that the gradient of the 3D MOT is shallow enough for its radial capture to extend beyond the cell bounds, and thus the integration limits are set from $-d_0$ to $d_0$, where $d_0$ is the distance a radial beam travels inside the cell to its center (3D MOT). In the second regime, the situation is opposite, and thus the integration limits are from $-d_\textsf{c}$ to $d_\textsf{c}$, where $d_\textsf{c}$ denotes the radial confinement distance. The integrations yield respective error functions $er\!f$ for the two regimes and, by mathematically separating them with appropriate Heaviside functions $H$, yield  
\begin{equation}
\label{eq:PO}
\hspace*{-85pt}
\begin{aligned}
\mathcal{P}_\oplus &=H(d_\textsf{c} - d_0) \times er\!f\left(\frac{d_0}{\sqrt{2\sigma_r^2}}\right) \\
                                &+H(d_0 - d_\textsf{c}) \times er\!f\left(\frac{d_\textsf{c}}{\sqrt{2\sigma_r^2}}\right)       
\end{aligned}
\end{equation}
\normalsize

\noindent where $H(0)=1/2$ for $d_\textsf{c}=d_0$ and continuity is preserved. The unknown variables are as follows:

\vspace*{5pt}
(i) $d_\textsf{c}=\frac{3|\delta_0|}{\mu B'}$, where $\frac{\delta_0}{2\pi}$ is the radial beam detuning, $\mu$ is the gyromagnetic ratio, and $B'>0$ is the axial gradient (twice stronger than the radial). This expression is obtained by assuming for simplicity that the furthest capture occurs midway between the peak capture distance $\frac{2|\delta_0|}{\mu B'}$ and twice this distance (where the radiation pressure force is significantly diminished). We note that in general, the confinement distance has a complex dependence on different trapping parameters \cite{3:1}.

(ii) $\sigma_r=H(\tau-t)\sigma_{r,t} + H(t-\tau)\sigma_{r,s}$, where the Heaviside functions separate the cases where the time $\tau$ until the metastable-state shelving is longer than the transfer time $t$ and vice versa; and $\sigma^2_{r,t}$ and $\sigma^2_{r,s}$ are the variances of the Gaussians describing the radial position spreads for

\leftskip=0.15cm\rightskip=-3cm 
\noindent these respective cases. When the shelving occurs, the corresponding speed Gaussian is assumed to freeze at $\tau$ and evolve the position Gaussian until $t$ by being in quadrature with the position Gaussian at $\tau$ and through a correlation term; thus, $\sigma_{r,s}=\sqrt{\sigma^2_{r,\tau} + \sigma^2_{v,\tau}(t-\tau)^2 + 2\rho\sigma_{r,\tau}\sigma_{v,\tau}(t-\tau)}$, where $\sigma^2_{r,\tau}$ and $\sigma^2_{v,\tau}$ are the variances of respectively the position and speed Gaussians at $\tau$, and $\rho=\frac{\sqrt{3}}{2}$ is the correlation between them (Brownian in nature). The remaining quantities seen in $\sigma_r$ are expressed as follows: The transfer time $t=\frac{d-d_\textsf{e}}{-v_{max}}>0$, where $d$ is the distance between the 2D and 3D MOTs; $d_\textsf{e}=\frac{\delta_\text{b} - \delta_\text{t}}{2\mu B'}$ is the axial equilibrium position of the 3D MOT, as we assume the bottom and top beams of respective detunings $\frac{\delta_\text{b}}{2\pi}$ and $\frac{\delta_\text{t}}{2\pi}$ have the same intensity; and $v_{max}=\frac{\delta_\text{b} - \delta_\text{t}}{2k}$ is the maximal speed achieved during the transfer, assuming there is no magnetic field in the transfer region ($k=\frac{2\pi}{\lambda}$ is the angular wavenumber of the blue transition). The time until shelving, $\tau$, is obtained using Eq. \hyperref[eq:tau_simple]{B1} with $\mathcal{E}=1$ (no repumping) and with $\Gamma_{23}'$ as in Eq. \hyperref[eq:Gamma_23]{B2}, but here we use a two-level model total saturation parameter for the bottom and top beams, $s_\text{bt} = \frac{I_\text{bt}/I_{sat}}{1+4\frac{(\delta_\text{b}-kv_{max})^2}{\Gamma^2}} + \frac{I_\text{bt}/I_{sat}}{1+4\frac{(\delta_\text{t}+kv_{max})^2}{\Gamma^2}}$ ($I_\text{bt}$ is the intensity of an axial beam). The variance $\sigma^2_{r,\tau}=\frac{2D_r\tau^3}{3^2m^2}$ ($\sigma^2_{r,t}$ uses $\tau\rightarrow t$) and $\sigma^2_{v,\tau}=\frac{2D_r\tau}{3m^2}$, where the transfer-region diffusion coefficient $D_r$ is obtained using Eq. \hyperref[eq:diff]{A13} with the total saturation parameter $s_\text{bt}$. Note, finally, that our definition of $\sigma_r$ assumes that the atomic beam originates from a single point. An improved assumption would be adding $\sigma_r$ in quadrature with the radial 2D MOT size, which is proportional to the square root of its temperature \cite{3:2}. We omit this here, as our main results (Sec. \hyperref[sec:MainResults_ii]{III.B}) are qualitatively unaffected by this change.

Next, let us consider writing the probability $\mathcal{P}_\ominus$ incorporating various 3D MOT cut offs as well as the change in the confinement strength as the MOT equilibrium position $d_\textsf{e}$ shifts with respect to the fixed radial beams center. The first cut off we introduce ensures that the MOT is empty when $\delta_\text{t}$ is positive and larger than $|\delta_\text{b}|$, as the sign of the confinement is then flipped. Our next two cut offs ensure the MOT is empty when the radial beams do not encompass $d_\textsf{e}$ (as set by the axial beam imbalance), given the locations $d_u$ and $d_l$ of the upper and lower radial beam edges with respect to the quadrupole field zero. Note that additional cut offs can be introduced to ensure an empty MOT when the transfer speed overcomes the axial capture speed and when $d_\textsf{e}$ exceeds the axial confinement distance; we omit these here as for our parameters the first cut off condition is stronger. We represent the introduced cut offs (three in total) with\;\;Heaviside functions.\;\;For the change in the\;\;confinement\;\;strength\;\;with

\leftskip=-3cm\rightskip=0.15cm 
\normalsize
\begin{figure*}
\vspace*{-353pt}
\hspace*{-145.5pt}\includegraphics[scale=0.97]{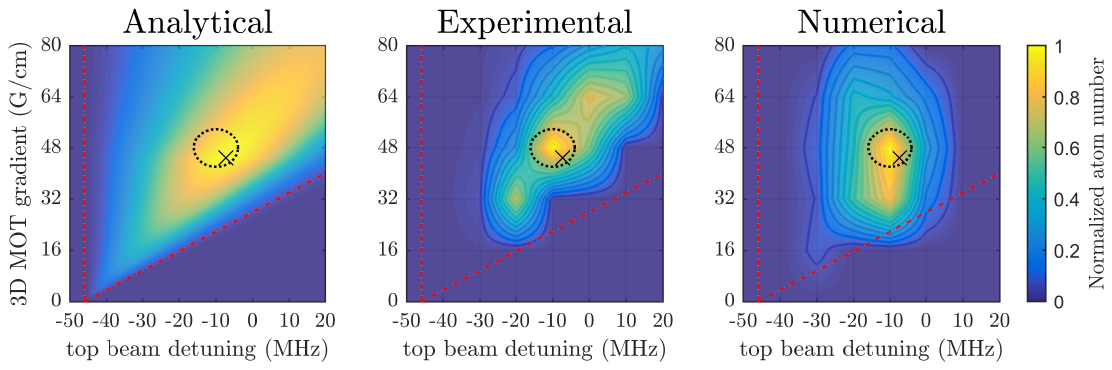}
\vspace*{-317pt}
 \captionsetup{width=1.49\linewidth}
  \caption*{\textbf{Figure 4:} Diagrams showing the 3D MOT atom numbers (normalized; see text for absolute) using our loading technique connecting the 2D and 3D MOTs via moving molasses, obtained analytically, experimentally, and numerically. The experimental and numerical data points are obtained at the values seen on the axes, with the contours enhancing the features. The vertical dashed line marks the fixed bottom beam detuning ($-46$ MHz), and the slanted dashed line delineates approximately the sloped analytical feature. The cross and the dotted circle highlight respectively the location with the greatest analytical atom number ($44.8$ G/cm gradient at $-7.3$ MHz top beam detuning) and the greatest experimental as well as numerical atom number ($48$ G/cm gradient at $-10$ MHz top beam detuning).}
\label{fig:4}
\end{figure*}

\noindent the shifting $d_\textsf{e}$, we assume a Gaussian behavior that depends on the trapped atom position $d_p$ with respect to the center of the radial beams ($w_0$ waist radius), where their intensity maximum is assumed to provide the strongest (unity) confinement. Thus,
\vspace*{-20pt}
\subsection*{}\label{test}
 
\begin{equation}
\label{eq:PX}
\hspace*{-95pt}
\hspace*{33pt}
\begin{aligned}
\mathcal{P}_\ominus &= H(-\delta_\text{t}-\delta_\text{b})H(d_{u}-d_\textsf{e})H(-d_l+d_\textsf{e})
\\ &\times exp\left(-2\frac{d_p^2}{w_0^2}\right) 
\end{aligned}
\end{equation}
The unknown variables in this equation are as follows:

(i) $d_u = d_a + d_{\text{off}}$ and $d_l = -d_a + d_{\text{off}}$, where $d_a$ is the optics aperture radius, and $d_{\text{off}}$ is the beam offset with respect to the quadrupole field zero.

(ii) $d_p=d_\textsf{e}-d_{\text{off}}$. 

For the final component in our analytical theory, we consider the time $\tau_\text{loss}$ until the dark-state decay occurs in the trap. We obtain it using Eq. \hyperref[eq:tau_simple]{B1} with $\mathcal{E}=27$ (repumping present) and with $\Gamma_{23}'$ as in Eq. \hyperref[eq:Gamma_23]{B2}, but here we use a two-level model total saturation parameter for all the 3D MOT beams, $s_\text{all}=s_\text{bt,B}+4s_0$, where $s_\text{bt,B}$ is the total saturation parameter given by $s_\text{bt}$ with the substitution $kv_{max}\rightarrow\mu B' d_\textsf{e}$ (assuming a motionless atom), and $s_0=\frac{\frac{I_{0}}{I_{sat}}exp\left(-2\frac{d_p^2}{w_0^2}\right)}{1+4\frac{\delta_0^2}{\Gamma^2}}$ is the radial beam saturation parameter, with $I_0$ the peak radial beam intensity. 

Putting all of the above together, we obtain the following equation for the atom number in the 3D MOT (when the transfer from the 2D MOT location occurs with the flow rate $\mathcal{F}_0$):

\begin{equation}
\label{eq:N}
\hspace*{-100pt}
\begin{aligned}
N = \frac{\mathcal{P}}{\Gamma_\text{loss}}\times\mathcal{F}_0 
\end{aligned}
\end{equation}
where $\mathcal{P}=\mathcal{P}_\oplus\times \mathcal{P}_\ominus$ combines the two probabilities $\mathcal{P}_\oplus$ and $\mathcal{P}_\ominus$\;\;\;affecting\;\;\;the overall\;\;\;transfer\;\;\;efficiency,\;\;\;and 

\leftskip=0.15cm\rightskip=-3cm
\noindent $\Gamma_\text{loss}=1/\tau_\text{loss}$ sets the dominant loss rate in the 3D MOT due to the dark-state decay. We note that $\Gamma_\text{loss}$ bounds the parameter space for optimal transfer, as discussed below.

\leftskip=-1.1cm
\subsection*{\hspace{1.61cm}\mbox{B. The experiment and comparison to}\\ \hspace*{2.1cm}\mbox{the analytical and numerical models}}\label{sec:MainResults_ii}
\leftskip=0.15cm\rightskip=-3cm
 
In Fig. \hyperref[fig:4]{4}, we compare the 3D MOT loading diagrams obtained analytically, experimentally, and numerically. The parameter space spans over different field gradients and top beam detunings. The shapes enveloping significant atom numbers are observed to take up roughly similar areas in all three cases, with the highest atom numbers (absolute values discussed below) obtained at the same data point locations in the experiment and simulation (dotted circle center, $48$ G/cm gradient at $-10$ MHz top beam detuning), while the analytical model prediction for the highest number is within the point's neighborhood (cross, $44.8$ G/cm gradient at $-7.3$ MHz top beam detuning). From our analytical model, we find that its optimal atom number prediction occurs for the absolute transfer speed (from the 2D MOT) $|v_{max}| \approx$ 9 m/s, which leads to the transfer time $t \approx$ 36 ms. This time is approximately five times longer than the time $\tau$ until the atoms go dark as they shelve to the metastable state $5s5p\,^3\!P_2$ during their transfer to the 3D MOT, where they subsequently get revived due to the repumping in this region (Sec. \hyperref[sec:Experiment_iv]{II.D}). Furthermore, we find the corresponding axial equilibrium position $d_\textsf{e} = -3$ mm (the minus sign refers to the position being below the 3D MOT quadrupole field zero), which is lower than the offset of the radial beams ($-2.4$ mm, with the absolute being the waist size; see Sec. \hyperref[sec:Experiment_iv]{II.D}). This difference is related to the atoms having an extended lifetime in the 3D MOT when not at the intensity maximum of the radial beams. 

\end{multicols}

\newgeometry{left=1.86in,right=1.86in,top=1.73in,bottom=0.6in}
\begin{multicols}{2}\setlength{\columnsep}{2pt}
\leftskip=-3cm\rightskip=0.15cm

For the remaining shape differences in the diagrams of the experiment and simulation (more complex than analytical theory), this may be attributed to the simulation being sensitive to the 3D MOT magnetic field modeling, as we use an approximation to the quadrupole field created by the rectangular coils in our experiment (see Sec. \hyperref[sec:Experiment_iv]{II.D}). The most significant magnetic field difference would be far from the 3D MOT center and in the transfer region itself, altering the loading dynamics. 

We use the analytical model to explain the different features in the diagrams as follows: 

(i) For the dark area near the line where the axial beams are equally detuned ($-46$ MHz; vertical dashed line), three parts are involved. Just to the left of this line, the atom transfer occurs in the opposite direction (larger top beam detuning). At this line, the transfer is obviously halted. Just to the right, the transfer is slow enough so the atoms disperse in significant numbers before reaching the 3D MOT region, with the higher gradients resulting in the dark area extending further towards resonance as the radial capture distance $d_\textsf{c}$ becomes smaller. 

(ii) The dark area on the right side of the diagrams (delineated by the slanted dashed line) is, on the other hand, obtained due to the axial equilibrium position $d_\textsf{e}$ being far enough below the radial 3D MOT beams center so the confinement is significantly diminished, or even non existent due to the radial beams not encompassing $d_\textsf{e}$. The boundary appears linear as $d_\textsf{e}$ is proportional to the axial beam detuning difference while inversely proportional to the gradient. 

(iii) Regarding the bright isle with the highest atom numbers, its location depends on the MOT confinement and the dark-state loss. The former expresses itself in the capture from the atomic\;\;beam (see the error functions in Eq. \hyperref[eq:PO]{1}) and the relative positioning of the radial 3D MOT beams (see the Gaussian in Eq. \hyperref[eq:PX]{2}), while the latter is encapsulated by the decay of the trap (see $\Gamma_\text{loss}$ in Eq. \hyperref[eq:N]{3}). We note that removing the decay results in the absence of a clear localization of the highest atom numbers, as shown in Fig. \hyperref[fig:AppB]{7} with an extended parameter space (compared to Fig. \hyperref[fig:4]{4}). The localization must otherwise be present as the trap lifetime becomes shorter for larger axial beam detuning imbalances and higher gradients in this considered space. We are therefore provided evidence that the dark-state loss is of fundamental importance for our observations.

We next compare the atom numbers. In the experiment, the highest atom number obtained is ${\sim}2.3\times10^5$ atoms with the loading rate of $1.5\times10^6$ atoms/s, which is expectedly bounded by that of the 2D MOT ($1.5\times10^9$ atoms/s). A same-order-of-magnitude value for the 3D MOT loading has been reported in the literature for a se- 

\leftskip=0.15cm\rightskip=-3cm 
\noindent tup employing a push beam for comparable experimental parameters \cite{42}. Nonetheless, there are adjustments to be made to greatly increase our loading rate, and, in Sec. \hyperref[sec:MainResults_iii]{III.C}, we discuss the numbers that can be attained. By using the experimental flow rate in the analytical model's Eq. \hyperref[eq:N]{3} ($\mathcal{P}\times\mathcal{F}_0=1.5\times10^6$ atoms/s), we obtain ${\sim}1.6\times10^5$ atoms at the peak location (cross in Fig. \hyperref[fig:4]{4}), which is in close agreement with the experiment (by $30\,\%$) and telling that our lifetime estimate reflects the reality well. Regarding the simulation, it predicts the maximum of ${\sim}3\times10^4$ atoms, which is roughly 8 times less than in the experiment. This discrepancy is in small part due to the 2D MOT loading rate discrepancy ($13\,\%$; see Tab. \hyperref[tab:1]{1}) but mostly due to the loading being diminished by the choice of the radial beams offset (1 waist below the quadrupole field zero), as we report a simulated ${\sim}2.5$ times enhancement when a smaller offset is used (around a third of a beam waist), giving thus an overall ${\sim}3$ times difference between the simulation and experiment. This numerical atom number dependence on the offset is, however, different from the analytical result discussed below.

\vspace*{-140pt}
\hspace*{-80pt}\includegraphics[scale=0.64]{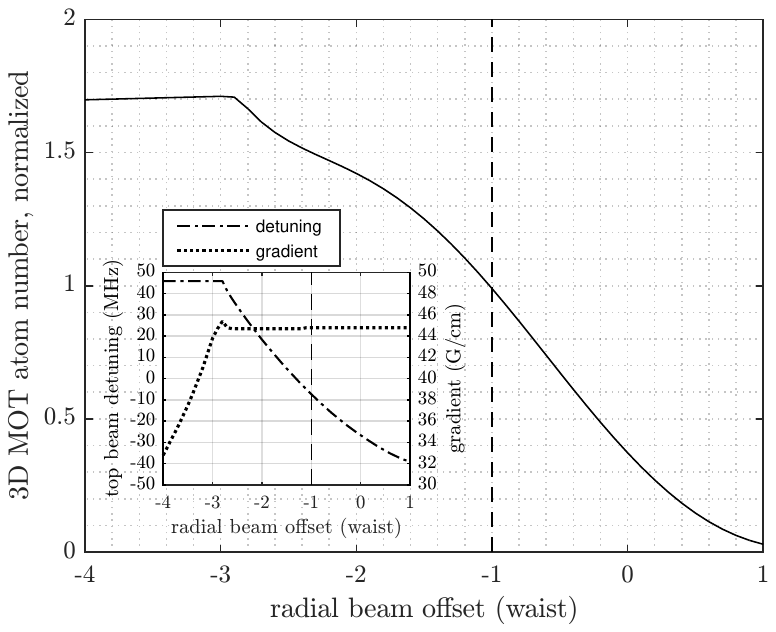}\label{fig:5}
\vspace*{-155pt}
\begin{footnotesize}
\begin{spacing}{1}
\vspace*{-3pt}\noindent{\textbf{Figure 5:} The maximal 3D MOT atom number dependence on the radial beam offset, obtained using our analytical model (Sec. \hyperref[sec:MainResults_i]{III.A}). The vertical dashed line indicates the offset used in the experiment (by 1 radial beam waist below the quadrupole field zero). The normalization is done with respect to the result at this offset. The inset shows the top beam detuning (dash-dotted line) and gradient (dotted line) dependence on this offset.\newline}
\end{spacing}
\end{footnotesize}

\leftskip=0.15cm\rightskip=-3cm 
\normalsize
\vspace*{25pt}
In Fig. \hyperref[fig:5]{5}, we provide an analytical calculation of the maximal 3D MOT atom number dependence on the radial beam offset. We first observe that by using the experimental beam offset (indicated by the vertical dashed line)

\leftskip=-3cm\rightskip=0.15cm
\noindent the atom number is seen to more than double compared to when no offset is used. Moreover, we observe that the atom number can be increased further by up to ${\sim}70\,\%$, given the beams are lowered by an additional 2 radial beam waists and a larger axial beam detuning imbalance is used (see the inset) for a more downshifted $d_\textsf{e}$. Note that an approximate plateau is observed beyond this offset as the limit for the top beam detuning is reached (refer to Fig. \hyperref[fig:AppB]{7}), at which point the gradient has to decrease such that $d_\textsf{e}$ can continue to lower. We expect the MOT to be empty for relatively large beam offsets, as the quadrupole MOT field needed for the confinement does not extend indefinitely, which is not considered in our analytical model. This detail is captured by our numerical model, which together with the included polarization effects can reasonably explain why the atom number should peak and eventually fall with increasing offset.

\leftskip=-4.58cm
\subsection*{\hspace{1.61cm}\mbox{C. The influence of metastable-state shel-}\\ \hspace*{1.9cm}\mbox{ving and misalignment on atom transfer}}\label{sec:MainResults_iii}
\leftskip=-3cm\rightskip=0.15cm
\normalsize

As implied by our analytical model (Sec. \hyperref[sec:MainResults_i]{III.A}), the 3D MOT loading efficiency critically depends on the dispersion the atoms accumulate (due to diffusion) during their transfer from the 2D MOT. This dispersion is statistically diminished by the metastable-state shelving (refer to the definition of $\sigma_r$ in Sec. \hyperref[sec:MainResults_i]{III.A}) but cannot be completely eliminated. It is thus natural to seek a configuration where the transfer speed is increased and the dispersion is kept low. We experimentally demonstrate this by misaligning the bottom beam (see Fig. \hyperref[fig:1]{1}), leading to a radiation-pressure imbalance without enhanced scattering. Using such a configuration, we report observing an approximately 2 times increase in the experimental atom number (to ${\sim}4.6\times10^5$ atoms). For this, the bottom beam misalignment of ${\sim}50$ mrad at the MOT center is used, with higher offset values resulting in an empty MOT. Moreover, we report that the radiation-pressure imbalance achieved using this special configuration makes it possible to trap with equally detuned axial beams, as our simulations also confirm. Note that, in the further discussion, we focus only on the misaligned case having the same MOT parameters (including the axial beam de-

\leftskip=0.15cm\rightskip=-3cm
\noindent tuning imbalance) as in the aligned case.

\vspace*{-150pt}
\hspace*{-80pt}\includegraphics[scale=0.64]{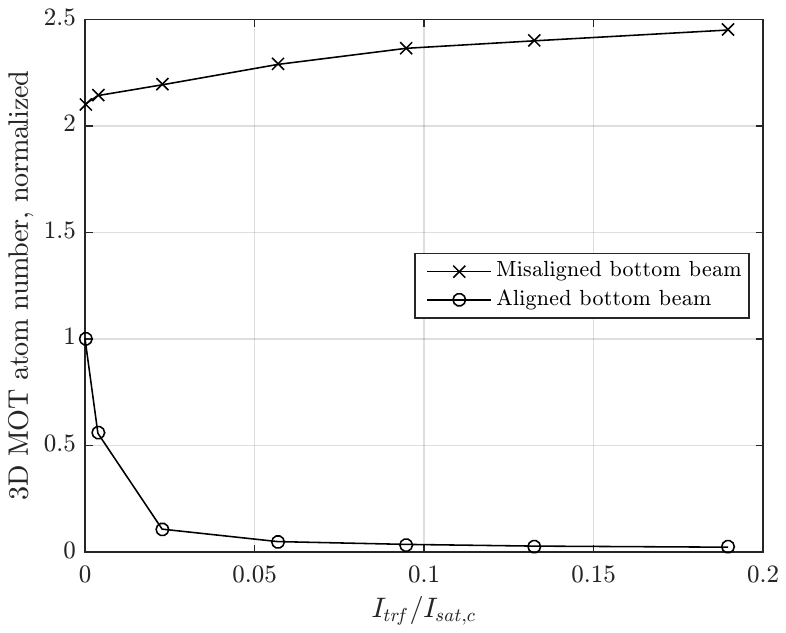}\label{fig:6}
\vspace*{-155pt}
\begin{footnotesize}
\begin{spacing}{1}
\vspace*{-3pt}\noindent{\textbf{Figure 6:} Experimental data showing the 3D MOT atom numbers (normalized) for the cases where the bottom beam is misaligned (crosses) and aligned (circles) vs the repumping intensity $I_{tr\!f}$ in the transfer region through the top beam (scaling with the saturation intensity $I_{sat,c}$). The misalignment is ${\sim}50$ mrad at the MOT center, and the remaining parameters are as in Fig. \hyperref[fig:4]{4} for the maximal atom number. The normalization is done with respect to the aligned case for zero transfer-region repumping intensity. Note that the data are obtained in all cases with a constant intensity of repumping light present in a single retroreflected radial beam pair (see text). \newline}
\end{spacing}
\end{footnotesize}

The misaligned configuration also opens a possibility of introducing repumping light in the transfer region, which is otherwise detrimental in the aligned configuration, as Fig. \hyperref[fig:6]{6}  implies. Here, we show experimentally obtained atom numbers in the 3D MOT for these respective cases (crosses and circles) versus the repumping intensity $I_{tr\!f}$ in the transfer region (overlapped with the top beam). We note that, for these measurements, repumping light of constant intensity is present in a retroreflected radial beam pair as in our original setting (see Sec. \hyperref[sec:Experiment_iv]{II.D}). The respective increase and decrease versus $I_{tr\!f}$ serves as an experimental proof that the metastable-state shelving (to $5s5p\,^3\!P_2$) is essential for the 3D MOT loading dynamics. Indeed, with the misalignment the speed of the atoms is increased and results in a rapid transfer with less time to disperse outside the MOT capture, whereas for no misalignment the atoms are relatively slow and disperse in significant numbers before reaching the MOT region due to heating. Adding repumping to the transfer region extends the time the atoms are affected by the blue light, creating (i) for the misaligned case, further acceleration downwards in the region where the top and bottom beams do not overlap, and (ii) for the aligned case, further detrimental heating leading to radial\;\,dispersion. With\;\;larger

\leftskip=-3cm\rightskip=0.15cm
\noindent intensities, the atomic medium becomes increasingly saturated, and hence the trends in Fig. \hyperref[fig:6]{6} for the misaligned and aligned cases follow. Note that this figure implies that we can achieve an atom number improvement of ${\sim}2.5$ compared to the aligned configuration, for the former case using a relatively low repumping intensity in the transfer region ($I_{tr\!f}/I_{sat,c}\approx0.2$). We can possibly enhance this number further by adding repumping light to the second pair of retroreflected radial beams, as the capture area for shelved atoms would be increased.

There exist adjustments in our system that would add significantly to the loading enhancement. One of the adjustments concerns the oven temperature (Sec. \hyperref[sec:Experiment_i]{II.A}), while the other concerns the ZS performance (Tab. \hyperref[tab:1]{1}). Our oven temperature of $440\,^{\circ}$C is relatively modest compared to the temperature of, e.g., $550\,^{\circ}$C reported in the literature \cite{36,39}; assuming that the capturable flux increases as $P_\text{Sr}(T)\times T^{-2}$ [$P_\text{Sr}(T)$ is a semi empirical equation for vapor-pressure of strontium at temperature $T$ in Kelvin \cite{3:3extra1,3:3extra2}, and $T^{-2}$ is the scaling of the 1D Maxwell-Boltzmann distribution in the low-speed limit], we expect around 25 times loading enhancement. By optimizing the ZS stage, we expect to bring the enhancement to two orders of magnitude \cite{36,39}; installing servomotors for the optimization may be a good practical choice \cite{3:3}, although we note that it is costly compared to using a numerical algorithm \cite{39,2:0}. All in all, by increasing the oven temperature to $550\,^{\circ}$C, optimizing the ZS stage, and working in the misaligned configuration while employing repumping considerations above, we expect the 3D MOT loading rate to exceed $4\times10^8$ atoms/s and be within an order of magnitude with the best performing Sr loading experiments \cite{35,36,39,3:4}.

\leftskip=-0.5cm
\section{\hspace{-0.38cm}Conclusions}\label{sec:Conclusions}
\leftskip=-3cm\rightskip=0.15cm

This paper describes and analyzes a setup integrating a moving molasses technique, where a 3D MOT is loaded using its axial beams creating a cold beam of atoms from the location of a 2D MOT (Fig. \hyperref[fig:1]{1}). The highest loading is achieved by misaligning the bottom axial beam with a different detuning than the top beam (by ${\sim}1\,\Gamma$). Loading rates exceeding $4\times10^8$ atoms/s are achievable with a high oven temperature ($550\,^{\circ}$C) and an optimized ZS, being well sufficient for obtaining dense red MOTs and large Sr tweezer arrays \cite{14,21,2:3,conc:0}. This technique completely excludes an additional push beam and in-vacuum parts, while resulting in the 3D MOT being loaded in an advantageous laser beam configuration. Moreover, it provides a broad optical access to the scientific region of study, allowing for the realization of high\;\;NA\;\;optical\;\;tweezers

\leftskip=0.15cm\rightskip=-3cm 
\noindent and opening more laser pathways for atom array manipulation, as highly desired  in Rydberg-atom based simulation and computing \cite{16,17,20,21}. 

We find the metastable-state shelving and the dark-state loss to be essential for the MOT loading dynamics. Particularly for the dark-state loss, it is shown by our analytical model to explain the existence of a well-localized parameter isle for highest atom numbers. This prediction is confirmed by our numerical model, which is more advanced than the analytical one (four-level versus two-level model). The corresponding simulation tool is used to verify the loading to high accuracy and can be employed in the design and optimization of effusive oven experiments. Indeed, this tool simulates the full 3D atom dynamics starting at the hot source and ending at the 3D MOT (with the only free parameter being the atomic flow rate measured after the oven exit) and allows for arbitrary beam geometries while taking into account the effects of the magnetic fields and the Poissonian atom decay.

As the final remark, we report employing our moving molasses technique in a separate experiment with alkali atoms (cesium) \cite{conc:1} to load large tweezer arrays, thus demonstrating that this technique can be extended beyond alkaline earth metals.

\leftskip=1.95cm
\section{\hspace{-0.38cm}Acknowledgments}\label{sec:Acknowledgments}
\leftskip=0.15cm\rightskip=-3cm 

We are grateful for contributions to the experimental setup by A. Rybicki and J. McDowell, and our cesium team for sharing their data \cite{conc:1}. This work was supported by an award from the W. M. Keck Foundation. 

\vspace*{40pt}
\par\hspace{56pt}\rule{4cm}{0.6pt}
\hspace{-88pt}\rule{2cm}{2pt}
\hspace{-74pt}\rule{3cm}{1.1pt}
\begingroup
\renewcommand{\section}[2]{}

\let\OLDthebibliography\thebibliography
\renewcommand\thebibliography[1]{
  \OLDthebibliography{#1}
  \setlength{\parskip}{4pt}
  \setlength{\itemsep}{1pt plus 0.3ex}
}

\endgroup

\newpage
\leftskip=0.5cm
\section*{\hspace{-0.42cm}Appendix A}\label{sec:Appendix_A1}
\leftskip=-1.1cm
\leftskip=-3cm\rightskip=0.15cm
This Appendix describes the radiation pressure effects in our 3D numerical model. Ref. \cite{2:7} is closely followed.

We base our model on the hyperfine transition $F=0 \rightarrow F'=1$, faithful to the blue (or red) trapping descriptions of $^{88}$Sr. Here, each of the three Zeeman transitions $m_0 = 0 \rightarrow m_1 = -1, 0,+1$ between the hyperfine levels is treated as an independent two-level system and  driven by, respectively, $\sigma^-$, $\pi$, $\sigma^+$ polarized light (refer to Fig. 1(b) in Ref. \cite{2:7}). This approximation is expected to hold only in the regime of low saturation, $s=\frac{I_{L}/I_{sat}}{1+\frac{4\Delta^2}{\Gamma^2}}\ll1$, where $I_{L}$ is the laser intensity, $I_{sat}$ is the saturation intensity, $\Delta=\omega_L - \omega_0$ is the (angular) detuning of the laser frequency $\omega_L$ from the atomic transition $m_0 = 0 \rightarrow m_1 = 0$ frequency $\omega_0$, and $\Gamma$ is the natural linewidth of the hyperfine transition.

We take into account arbitrary beam directions when describing the following two physicals effects: (i) the mean radiation pressure force stemming from a beam, hereafter referred to, for brevity, as the radiation pressure force, and (ii) the diffusion resulting from its fluctuations. We discuss these effects for the case of a 2D optical molasses (2D OM), a Zeeman slower (ZS), a 2D magneto-optical trap (2D MOT), and a 3D MOT. The corresponding numerical setup is seen in Fig. \hyperref[fig:2]{2}(b), for reference.

Note that the multiple-scattering effects (attenuation and rescattering) described in Ref. \cite{2:7} can be straightforwardly included in this model.

\leftskip=-2.0cm
\subsection{\hspace{-0.cm}Radiation pressure force}\label{sec:A_i}

\leftskip=-3cm\rightskip=0.15cm 
To describe the radiation pressure force, we use the standard Doppler model ($s\ll1$ holds). For the complete system that includes the 3D MOT, 2D MOT, 2D OM, and ZS, this force reads as

\begin{equation}
\label{eq:F_tot}
\tag{A1}
\begin{aligned}
\hspace*{-89.5pt}
\textbf{F}_{tot}(\textbf{r},\textbf{v}) = \textbf{F}_{\textsf{3M}}(\textbf{r},\textbf{v}) + \textbf{F}_{\textsf{2M}}(\textbf{r},\textbf{v}) + \textbf{F}_{\textsf{2O}}(\textbf{r},\textbf{v}) + \textbf{F}_{\textsf{ZS}}(\textbf{r},\textbf{v})
\end{aligned}
\end{equation}
with the respective forces being
\begin{equation}
\label{eq:F_tot_components}
\tag{A2}
\begin{aligned}
\hspace*{-92pt}&\textbf{F}_{\textsf{3M}}(\textbf{r},\textbf{v})&&= \sum_{\substack{q = \sigma^+,\sigma^-,\pi \\ j = +,-}}\textbf{F}_{\textsf{x}',q;\textsf{3M}}^{j,\sigma^-} + \textbf{F}_{\textsf{y}',q;\textsf{3M}}^{j,\sigma^-} + \textbf{F}_{\textsf{z}',q;\textsf{3M}}^{j,\sigma^+}\\
\hspace*{-92pt}&\textbf{F}_{\textsf{2M}}(\textbf{r},\textbf{v})&&= \sum_{\substack{q = \sigma^+,\sigma^-,\pi \\ j = +,-}}\textbf{F}_{\textsf{x}',q;\textsf{2M}}^{j,\sigma^-} + \textbf{F}_{\textsf{y}',q;\textsf{2M}}^{j,\sigma^+}\\
\hspace*{-92pt}&\textbf{F}_{\textsf{2O}}(\textbf{r},\textbf{v})&&= \sum_{\substack{q = \sigma^+,\sigma^-,\pi \\ j = +,-}}\textbf{F}_{\textsf{y}',q;\textsf{2O}}^{j,\pi} + \textbf{F}_{\textsf{z}',q;\textsf{2O}}^{j,\pi}\\
\hspace*{-92pt}&\textbf{F}_{\textsf{ZS}}(\textbf{r},\textbf{v})&&= \sum_{q = \sigma^+,\sigma^-,\pi}\textbf{F}_{\textsf{x}',q;\textsf{ZS}}^{-,\pi}
\end{aligned}
\end{equation}

\leftskip=0.15cm\rightskip=-3cm
\noindent where the individual force components [in positive ($j=+$) and negative ($j=-$) axis directions] are expressed by
\begin{equation}
\label{eq:F_1}
\tag{A3}
\begin{aligned}
\hspace*{22pt}
\textbf{F}_{\alpha',q;\mathbbm{1}}^{\pm,Q}(\textbf{r},\textbf{v}) = \pm\frac{p_{\alpha',q;\mathbbm{1}}^{\pm,Q}(\textbf{r}){I}_{\alpha';\mathbbm{1}}^{\pm}(\textbf{r})\sigma_{\alpha',q;\mathbbm{1}}^{\pm}(\textbf{r},\textbf{v}) }{c}\hat{\boldsymbol{\alpha}}'_\mathbbm{1}
\end{aligned}
\end{equation}
In this equation, $\mathbbm{1}=\textsf{3M},\textsf{2M},\textsf{2O}$, or $\textsf{ZS}$ (respectively referring to the 3D MOT, 2D MOT, 2D OM, or ZS); $c$ is the vacuum light speed; $\textbf{r}=(x,y,z)$ is the atom position; $\textbf{v}=(v_\mathsf{x},v_\mathsf{y},v_\mathsf{z})$ is the atom velocity; $q$ refers to the respective $\sigma^-$, $\pi$, $\sigma^+$ atomic transitions, with $Q$ referring to the corresponding beam polarizations; and the remaining quantities are defined as follows. 

The unit vector $\pm\hat{\boldsymbol{\alpha}}'_\mathbbm{1}$ denotes an arbitrary direction (for 2D MOT and 2D OM only two of the three subequations below are applicable, and for ZS only one):
\makeatletter
    \def\tagform@#1{\maketag@@@{\normalsize(#1)\@@italiccorr}}
\makeatother
\normalsize
\begin{equation}
\label{eq:wavevectors}
\tag{A4}
\hspace*{3pt}
\begin{aligned}
\pm\hat{\boldsymbol{\mathsf{x}}}'_\mathbbm{1} &= \pm (\text{cos}(\phi_{\mathsf{x}';\mathbbm{1}}^{\pm,\mathsf{y}})\text{cos}(\phi_{\mathsf{x}';\mathbbm{1}}^{\pm,\mathsf{z}}),\text{sin}(\phi_{\mathsf{x}';\mathbbm{1}}^{\pm,\mathsf{z}}),-\text{sin}(\phi_{\mathsf{x}';\mathbbm{1}}^{\pm,\mathsf{y}})\text{cos}(\phi_{\mathsf{x}';\mathbbm{1}}^{\pm,\mathsf{z}}))\\
\hspace*{3pt}\pm\hat{\boldsymbol{\mathsf{y}}}'_\mathbbm{1} &= \pm (-\text{sin}(\phi_{\mathsf{y}';\mathbbm{1}}^{\pm,\mathsf{z}}),\text{cos}(\phi_{\mathsf{y}';\mathbbm{1}}^{\pm,\mathsf{x}})\text{cos}(\phi_{\mathsf{y}';\mathbbm{1}}^{\pm,\mathsf{z}}),\text{sin}(\phi_{\mathsf{y}';\mathbbm{1}}^{\pm,\mathsf{x}})\text{cos}(\phi_{\mathsf{y}';\mathbbm{1}}^{\pm,\mathsf{z}}))\\
\hspace*{3pt}\pm\hat{\boldsymbol{\mathsf{z}}}'_\mathbbm{1} &= \pm (\text{sin}(\phi_{\mathsf{z}';\mathbbm{1}}^{\pm,\mathsf{y}}),-\text{sin}(\phi_{\mathsf{z}';\mathbbm{1}}^{\pm,\mathsf{x}})\text{cos}(\phi_{\mathsf{z}';\mathbbm{1}}^{\pm,\mathsf{y}}),\text{cos}(\phi_{\mathsf{z}';\mathbbm{1}}^{\pm,\mathsf{x}})\text{cos}(\phi_{\mathsf{z}';\mathbbm{1}}^{\pm,\mathsf{y}}))
\end{aligned}
\end{equation}
\normalsize
\noindent where $\phi_{\alpha';\mathbbm{1}}^{\pm,\alpha}\in[0,2\pi)$ are the rotation angles around the $\alpha = \mathsf{x},\mathsf{y},\mathsf{z}$ axis for the positive or negative ($\pm$) $\alpha' = \mathsf{x}',\mathsf{y}',\mathsf{z}'$ (rotated axis) beam.

The coefficient $p_{\alpha',q;\mathbbm{1}}^{\pm,Q}$ denotes the fraction of the $\pm{\boldsymbol{\alpha}}'_\mathbbm{1}$ directed light of $\sigma^-$, $\pi$, or $\sigma^+$ polarization (described by $Q$) that drives the corresponding transition (described by $q$):
\begin{equation}
\label{eq:frac_p}
\tag{A5}
\hspace*{11pt}
\begin{aligned}
 &p_{\alpha',q;\mathbbm{1}}^{\pm,\sigma^+}(\textbf{r})&&=\begin{cases}
               \left( \frac{1}{2}\left[ 1 + \frac{\textbf{k}_{\alpha';\mathbbm{1}}^{\pm}\cdot\textbf{B}(\textbf{r})}{|\textbf{k}_{\alpha';\mathbbm{1}}^{\pm}||\textbf{B}(\textbf{r})|} \right] \right)^2\quad\quad\;\;,\quad q = \sigma^+\\
               \left( \frac{1}{2}\left[ 1 - \frac{\textbf{k}_{\alpha';\mathbbm{1}}^{\pm}\cdot\textbf{B}(\textbf{r})}{|\textbf{k}_{\alpha';\mathbbm{1}}^{\pm}||\textbf{B}(\textbf{r})|} \right] \right)^2\quad\quad\;\;,\quad q = \sigma^-\\
               1-(p_{\alpha',\sigma^+;\mathbbm{1}}^{\pm,\sigma+}+p_{\alpha',\sigma^-;\mathbbm{1}}^{\pm,\sigma+})\;\;\;\,\quad\;\,,\quad q = \pi
            \end{cases}\\
 &p_{\alpha',q;\mathbbm{1}}^{\pm,\sigma^-}(\textbf{r})&&=\begin{cases}
               \left( \frac{1}{2}\left[ 1 - \frac{\textbf{k}_{\alpha';\mathbbm{1}}^{\pm}\cdot\textbf{B}(\textbf{r})}{|\textbf{k}_{\alpha';\mathbbm{1}}^{\pm}||\textbf{B}(\textbf{r})|} \right] \right)^2\quad\quad\;\;,\quad q = \sigma^+\\
               \left( \frac{1}{2}\left[ 1 + \frac{\textbf{k}_{\alpha';\mathbbm{1}}^{\pm}\cdot\textbf{B}(\textbf{r})}{|\textbf{k}_{\alpha';\mathbbm{1}}^{\pm}||\textbf{B}(\textbf{r})|} \right] \right)^2\quad\quad\;\;,\quad q = \sigma^-\\
               1-(p_{\alpha',\sigma^+;\mathbbm{1}}^{\pm,\sigma-}+p_{\alpha',\sigma^-;\mathbbm{1}}^{\pm,\sigma-})\;\;\;\;\quad\,\,,\quad q = \pi
            \end{cases}\\
 &p_{\alpha',q;\mathbbm{1}}^{\pm,\pi}(\textbf{r})&&=\begin{cases}
               \frac{1}{2}\left(1-p_{\alpha',\pi;\mathbbm{1}}^{\pm,\pi}\right)\quad\quad\quad\quad\quad\quad\;,\quad q = \sigma^+\\
               \frac{1}{2}\left(1-p_{\alpha',\pi;\mathbbm{1}}^{\pm,\pi}\right)\quad\quad\quad\quad\quad\quad\;,\quad q = \sigma^-\\
               \left(\frac{\textbf{k}_{\alpha';\mathbbm{1}}^{\pm}\cdot\textbf{B}(\textbf{r})}{|\textbf{k}_{\alpha';\mathbbm{1}}^{\pm}||\textbf{B}(\textbf{r})|}\right)^2\quad\quad\quad\quad\quad\quad\,,\quad q = \pi
            \end{cases}
\end{aligned}
\end{equation}
\noindent where $\textbf{k}_{\alpha';\mathbbm{1}}^{\pm} = \pm k_L\hat{\boldsymbol{\alpha}}'_\mathbbm{1}$ is the wavevector with the wavenumber $k_L=\omega_L/c$; and $\textbf{B}$ is the total magnetic field experienced by the atom, written as the sum of the magnetic fields stemming from the different parts of the system. Their respective components are written below.

\leftskip=-3cm\rightskip=0.15cm 
\indent (i) For the 3D MOT:

\makeatletter
    \def\tagform@#1{\maketag@@@{\normalsize(#1)\@@italiccorr}}
\makeatother
\begin{equation}
\label{eq:B_3DMOT}
\normalsize{\tag{A6}}
\small
\hspace*{-100pt}
\begin{aligned}
B_{\mathsf{x_\textsf{3M}}}&=\frac{B'R}{2}f\frac{x_\textsf{3M}}{r_\textsf{3M}} \times \left( \frac{1}{\left(1+[\frac{r_\textsf{3M}}{R}+\frac{\eta}{2}]^2\right)^{3/2}} - \frac{1}{\left(1+[\frac{r_\textsf{3M}}{R}-\frac{\eta}{2}]^2\right)^{3/2}} \right)\\
B_{\mathsf{y_\textsf{3M}}}&=\frac{B'R}{2}f\frac{y_\textsf{3M}}{r_\textsf{3M}} \times \left( \frac{1}{\left(1+[\frac{r_\textsf{3M}}{R}+\frac{\eta}{2}]^2\right)^{3/2}} - \frac{1}{\left(1+[\frac{r_\textsf{3M}}{R}-\frac{\eta}{2}]^2\right)^{3/2}} \right)\\
B_{\mathsf{z_\textsf{3M}}}&=-B'Rf \times \left( \frac{1}{\left(1+[\frac{z_\textsf{3M}}{R}+\frac{\eta}{2}]^2\right)^{3/2}} - \frac{1}{\left(1+[\frac{z_\textsf{3M}}{R}-\frac{\eta}{2}]^2\right)^{3/2}} \right)
\end{aligned}
\end{equation}
\normalsize

\noindent where $r_\mathsf{3M}=\sqrt{x_\mathsf{3M}^2+y_\mathsf{3M}^2+z_\mathsf{3M}^2}$ is the total radial distance with $x_\mathsf{3M},y_\mathsf{3M},z_\mathsf{3M}$ being the 3D MOT coordinates (respectively equal to $x,y,z$ minus the corresponding 3D MOT center coordinate); $B'$ is the field gradient; $R$ is the coil radius; $\eta$ is the ratio of the coil separation to the coil radius; and the coefficient $f=\frac{\left(1+\frac{\eta^2}{4}\right)^{5/2}}{3\eta}$. It is well known that the general anti-Helmholtz field involves elliptic integrals, and here a less complicated form of its radial and axial components is assumed. Note that we take into account the radial components vanishing for distances far from the trap center. Also note that the field approaches $-B'\left(\frac{x_\textsf{3M}}{2},\frac{y_\textsf{3M}}{2},-z_\textsf{3M}\right)$ when working on-axis and close to the trap center, usually considered in MOT descriptions.

\indent (ii) For the 2D MOT:

\begin{equation}
\label{eq:B_2DMOT}
\tag{A7}
\hspace*{-115pt}
\begin{aligned}
B_{\mathsf{x_\textsf{2M}}}&=B^{\wedge,+}\times(3y_\textsf{2M}^{\wedge,+}x_\textsf{2M}^{\wedge,+})\\
&+B^{\wedge,-}\times(-3y_\textsf{2M}^{\wedge,-}x_\textsf{2M}^{\wedge,-})\\
&+B^{\vee,+}\times(3y_\textsf{2M}^{\vee,+}x_\textsf{2M}^{\vee,+})\\
&+B^{\vee,-}\times(-3y_\textsf{2M}^{\vee,-}x_\textsf{2M}^{\vee,-})\\
B_{\mathsf{y_\textsf{2M}}}&=B^{\wedge,+}\times(2[y_\textsf{2M}^{\wedge,+}]^2 - [x_\textsf{2M}^{\wedge,+}]^2 - [z_\textsf{2M}^{\wedge,+}]^2)\\
&+B^{\wedge,-}\times(-2[y_\textsf{2M}^{\wedge,-}]^2 + [x_\textsf{2M}^{\wedge,-}]^2 + [z_\textsf{2M}^{\wedge,-}]^2)\\
&+B^{\vee,+}\times(2[y_\textsf{2M}^{\vee,+}]^2 - [x_\textsf{2M}^{\vee,+}]^2 - [z_\textsf{2M}^{\vee,+}]^2)\\
&+B^{\vee,-}\times(-2[y_\textsf{2M}^{\vee,-}]^2 + [x_\textsf{2M}^{\vee,-}]^2 + [z_\textsf{2M}^{\vee,-}]^2)\\
B_{\mathsf{z_\textsf{2M}}}&=B^{\wedge,+}\times(3y_\textsf{2M}^{\wedge,+}z_\textsf{2M}^{\wedge,+})\\
&+B^{\wedge,-}\times(-3y_\textsf{2M}^{\wedge,-}z_\textsf{2M}^{\wedge,-})\\
&+B^{\vee,+}\times(3y_\textsf{2M}^{\vee,+}z_\textsf{2M}^{\vee,+})\\
&+B^{\vee,-}\times(-3y_\textsf{2M}^{\vee,-}z_\textsf{2M}^{\vee,-})
\end{aligned}
\end{equation}
\noindent where ``$\wedge$" and ``$\vee$" in the superscripts refer to the magnets respectively above and below the center xy plane of the 2D MOT, while ``$+$" and ``$-$" refer to them being respectively positively and negatively displaced in the x direction with respect to the 2D MOT center. $B^{*}=\frac{B_0 V}{4\pi([x_\textsf{2M}^{*}]^2+[y_\textsf{2M}^{*}]^2+[z_\textsf{2M}^{*}]^2)^{5/2}}$ is the field magnitude of the corresponding magnet, with $B_0$ being the remanent magnetization and $V$ the magnet volume; and $x_\textsf{2M}^{*}=x_\textsf{2M}-x^{*}$, $y_\textsf{2M}^{*}=y_\textsf{2M}-y^{*}$, $z_\textsf{2M}^{*}=z_\textsf{2M}-z^{*}$, where $x_\textsf{2M},y_\textsf{2M},z_\textsf{2M}$ are 

\leftskip=0.15cm\rightskip=-3cm
\noindent the 2D MOT coordinates, and $x^{*},y^{*},z^{*}$ are the magnet coordinates (``*" refers to either ``$\wedge,+$", ``$\wedge,-$", ``$\vee,+$", or ``$\vee,-$"). We note that Eq. \ref{eq:B_2DMOT} has been obtained using the field equation for a magnetic dipole.

\indent (iii) For the 2D OM, the magnetic field is not present [i.e., $\textbf{0}=(0,0,0)$]. This will result in only the $\sigma^+$ and $\sigma^-$ transitions being driven according to Eq. \ref{eq:frac_p}. 

\indent (iv) For the ZS, we use a heuristic model based on segmentation with linear functions:
\begin{equation}
\label{eq:B_Zeeman}
\tag{A8}
\vspace*{-3pt}
\hspace*{5pt}
\begin{aligned}
B_{\mathsf{x_\textsf{ZS}}}&=0\\
B_{\mathsf{y_\textsf{ZS}}}&=
\sum_{n=1,2,...,n_{S}}\left\{H(x_\textsf{ZS}-(n-1)L/n_{S})-H(x_\textsf{ZS}-nL/n_{S})\right\}\\
&\times \left\{\left[(B_{n+1}-B_{n})/(L/n_S)\right]\times\left[x_\textsf{ZS}-(n-1)L/n_S\right] + B_{n}\right\}\\
B_{\mathsf{z_\textsf{ZS}}}&=0
\end{aligned}
\end{equation}

\noindent where the Heaviside functions $H$ provide a domain of truncation, with the outside field being zero; $x_\textsf{ZS}$ is the $\mathsf{x}$ coordinate of the ZS (obtained by subtracting $x$ with the corresponding ZS entrance coordinate); $L$ is the length of the ZS; $B_{1,2,...,n_S+1}$ is the size of the magnetic field at a given location, with $B_1$ and $B_{n_S+1}$ being the values at the ZS entrance and exit, respectively, where $n_S$ is the amount of segments. We note that the magnetic field here is perpendicular to the non rotated propagation direction of the Zeeman beam ($-\hat{\boldsymbol{\mathsf{x}}}$). For a two-level model, this would result in a vanishing radiation pressure force; on the other hand, as our model is sensitive to light polarization, the $\sigma^+$ and $\sigma^-$ transitions will be driven according to Eq. \ref{eq:frac_p}. 
\newline\newline\newline
The beam intensity ${I}_{\alpha';\mathbbm{1}}^{\pm}$ in Eq. \ref{eq:F_1} follows a truncated Gaussian profile (for 2D MOT and 2D OM only two of the three subequations below are applicable, and for ZS only one):
\begin{equation}
\label{eq:Intensity}
\tag{A9}
\hspace*{115pt}
\begin{aligned}
\hspace*{-110pt}{I}^{\pm}_{\mathsf{x}';\mathbbm{1}}(\textbf{r})&=C\left[y_{\mathsf{x}';\mathbbm{1}}^{\pm}(\textbf{r}),z_{\mathsf{x}';\mathbbm{1}}^{\pm}(\textbf{r}),h_{\mathsf{x};\mathbbm{1}}^{\pm}\right]\times{I}^{\pm}_{\mathsf{x},0;\mathbbm{1}}e^{-2\frac{\left(y_{\mathsf{x}';\mathbbm{1}}^{\pm}(\textbf{r})\right)^2 + \left(z_{\mathsf{x}';\mathbbm{1}}^{\pm}(\textbf{r})\right)^2}{\left(w^{\pm}_{\mathsf{x},0;\mathbbm{1}}\right)^2}}\\
\hspace*{-110pt}{I}^{\pm}_{\mathsf{y}';\mathbbm{1}}(\textbf{r})&=C\left[x_{\mathsf{y}';\mathbbm{1}}^{\pm}(\textbf{r}),z_{\mathsf{y}';\mathbbm{1}}^{\pm}(\textbf{r}),h_{\mathsf{y};\mathbbm{1}}^{\pm}\right]\times{I}^{\pm}_{\mathsf{y},0;\mathbbm{1}}e^{-2\frac{\left(x_{\mathsf{y}';\mathbbm{1}}^{\pm}(\textbf{r})\right)^2 + \left(z_{\mathsf{y}';\mathbbm{1}}^{\pm}(\textbf{r})\right)^2}{\left(w^{\pm}_{\mathsf{y},0;\mathbbm{1}}\right)^2}}\\
\hspace*{-110pt}{I}^{\pm}_{\mathsf{z}';\mathbbm{1}}(\textbf{r})&=C\left[x_{\mathsf{z}';\mathbbm{1}}^{\pm}(\textbf{r}),y_{\mathsf{z}';\mathbbm{1}}^{\pm}(\textbf{r}),h_{\mathsf{z};\mathbbm{1}}^{\pm}\right]\times{I}^{\pm}_{\mathsf{z},0;\mathbbm{1}}e^{-2\frac{\left(x_{\mathsf{z}';\mathbbm{1}}^{\pm}(\textbf{r})\right)^2 + \left(y_{\mathsf{z}';\mathbbm{1}}^{\pm}(\textbf{r})\right)^2}{\left(w^{\pm}_{\mathsf{z},0;\mathbbm{1}}\right)^2}}
\end{aligned}
\end{equation}

\noindent where ${I}^{\pm}_{\alpha,0;\mathbbm{1}}$ is the peak intensity of the positive or negative ($\pm$) $\alpha=\mathsf{x},\mathsf{y},\mathsf{z}$ beam, $w^{\pm}_{\alpha,0;\mathbbm{1}}$ is the corresponding waist radius; 

\leftskip=-3cm\rightskip=0.15cm
\small
\begin{equation}
\label{eq:Intensity_coordinates}
\tag{A10}
\hspace*{20pt}
\begin{aligned}
\hspace*{-113pt} y_{\mathsf{x}';\mathbbm{1}}^{\pm}(\textbf{r}) &= \mp\text{cos}(\phi_{\mathsf{x}';\mathbbm{1}}^{\pm,\mathsf{y}})\text{sin}(\phi_{\mathsf{x}';\mathbbm{1}}^{\pm,\mathsf{z}})x \pm \text{cos}(\phi_{\mathsf{x}';\mathbbm{1}}^{\pm,\mathsf{z}})y \pm \text{sin}(\phi_{\mathsf{x}';\mathbbm{1}}^{\pm,\mathsf{y}})\text{sin}(\phi_{\mathsf{x}';\mathbbm{1}}^{\pm,\mathsf{z}})z\\
\hspace*{-113pt} z_{\mathsf{x}';\mathbbm{1}}^{\pm}(\textbf{r}) &= \pm\text{sin}(\phi_{\mathsf{x}';\mathbbm{1}}^{\pm,\mathsf{y}})x \pm \text{cos}(\phi_{\mathsf{x}';\mathbbm{1}}^{\pm,\mathsf{y}})z\\
\hspace*{-113pt} x_{\mathsf{y}';\mathbbm{1}}^{\pm}(\textbf{r}) &= \pm\text{cos}(\phi_{\mathsf{y}';\mathbbm{1}}^{\pm,\mathsf{z}})x \pm \text{cos}(\phi_{\mathsf{y}';\mathbbm{1}}^{\pm,\mathsf{x}})\text{sin}(\phi_{\mathsf{y}';\mathbbm{1}}^{\pm,\mathsf{z}})y \pm \text{sin}(\phi_{\mathsf{y}';\mathbbm{1}}^{\pm,\mathsf{x}})\text{sin}(\phi_{\mathsf{y}';\mathbbm{1}}^{\pm,\mathsf{z}})z\\
\hspace*{-113pt} z_{\mathsf{y}';\mathbbm{1}}^{\pm}(\textbf{r}) &= \mp\text{sin}(\phi_{\mathsf{y}';\mathbbm{1}}^{\pm,\mathsf{x}})y \pm \text{cos}(\phi_{\mathsf{y}';\mathbbm{1}}^{\pm,\mathsf{x}})z\\
\hspace*{-113pt} x_{\mathsf{z}';\mathbbm{1}}^{\pm}(\textbf{r}) &= \pm\text{cos}(\phi_{\mathsf{z}';\mathbbm{1}}^{\pm,\mathsf{y}})x \pm \text{sin}(\phi_{\mathsf{z}';\mathbbm{1}}^{\pm,\mathsf{x}})\text{sin}(\phi_{\mathsf{z}';\mathbbm{1}}^{\pm,\mathsf{y}})y \mp \text{cos}(\phi_{\mathsf{z}';\mathbbm{1}}^{\pm,\mathsf{x}})\text{sin}(\phi_{\mathsf{z}';\mathbbm{1}}^{\pm,\mathsf{y}})z\\
\hspace*{-113pt} y_{\mathsf{z};\mathbbm{1}'}^{\pm}(\textbf{r}) &= \pm\text{cos}(\phi_{\mathsf{z}';\mathbbm{1}}^{\pm,\mathsf{x}})y \pm \text{sin}(\phi_{\mathsf{z}';\mathbbm{1}}^{\pm,\mathsf{x}})z
\end{aligned}
\end{equation}
\normalsize

\noindent are the rotated coordinates, with the offsets supressed for brevity (they can be reintroduced by performing appropriate subtractions to $x,y,z$); and $C[\cdot,:,h_{\alpha;\mathbbm{1}}^{\pm}]$ is the cylinder function with $h_{\alpha;\mathbbm{1}}^{\pm}$ being its cut off [$C=1$ for $\sqrt{(\cdot)^2+(:)^2}\leq h_{\alpha;\mathbbm{1}}^{\pm}$, and $0$ otherwise]. The truncation is naturally considered, as the beams pass through the optics apertures limiting their size.

The final quantity in Eq. \ref{eq:F_1}, $\sigma_{\alpha',q;\mathbbm{1}}^{\pm}$, is the corresponding scattering cross section for a single two-level atomic transition:

\begin{equation}
\label{eq:sigma}
\tag{A11}
\begin{aligned}
\hspace*{-90pt}
\sigma^{\pm}_{\alpha',q;\mathbbm{1}}(\textbf{r},\textbf{v})=\frac{\sigma_0}{1+\frac{I_{tot,q}(\textbf{r})}{I_{sat}}+4\frac{(\Delta_\mathbbm{1} - \textbf{k}_{\alpha';\mathbbm{1}}^{\pm}\cdot\textbf{v} - \mu_q(\textbf{r}))^2}{\Gamma^2}}
\end{aligned}
\end{equation}
\leftskip=-3cm\rightskip=0.15cm
\noindent where $\sigma_0=6\pi/k_L^2$ is the resonant scattering cross section; the detuning $\Delta_\mathbbm{1}=\Delta_{\textsf{3M}}$, $\Delta_{\textsf{2M}}$, $\Delta_{\textsf{2O}}$, or $\Delta_{\textsf{ZS}}$ is that of the 3D MOT, 2D MOT, 2D OM, or ZS beams, respectively; $\textbf{k}_{\alpha';\mathbbm{1}}^{\pm}\cdot\textbf{v}$ is the Doppler shift for a positive or negative ($\pm$) beam; and $\mu_q(\textbf{r})= q\mu B(\textbf{r})$ is the Zeeman shift for the $m_1 = -1, 0,+1$ level (where, respectively, $q=-,0,+$), with $\mu$ being the gyromagnetic ratio (the particular Zeeman shift is due to the quantization axis being chosen to be along the direction of $\textbf{B}$);

\begin{equation}
\label{eq:Intensity_tot}
\tag{A12}
\begin{aligned}
\hspace*{-92pt}
I_{tot,q}(\textbf{r})=\sum_\mathbbm{1}I_{tot,q;\mathbbm{1}}(\textbf{r},\textbf{v})
\end{aligned}
\end{equation}

\noindent is the total beam intensity that a single two-level transition receives, which is seen to be a sum of the corresponding total beam intensities of the 3D MOT, 2D MOT, 2D OM, and ZS, respectively. For the 3D MOT, for instance, one has $I_{tot,q;\textsf{3M}}=\sum_{\alpha'}p_{\alpha',q;\textsf{3M}}^{+,Q}(\textbf{r}){I}^{+}_{\alpha';\textsf{3M}}(\textbf{r}) + p_{\alpha',q;\textsf{3M}}^{-,Q}(\textbf{r}){I}^{-}_{\alpha';\textsf{3M}}(\textbf{r})$, where $Q=\sigma^-$ for $\alpha' = \mathsf{x}',\mathsf{y}'$ and $Q=\sigma^+$ for $\alpha'=\mathsf{z}'$.

\vspace*{15pt}
\leftskip=0.0cm
\subsection{\hspace{-0.cm}Diffusion}\label{sec:A_ii}
\leftskip=-3cm\rightskip=0.15cm

The fluctuating part of the trapping force is introduced here via a momentum diffusion coefficient that we write as follows:

\leftskip=0.15cm\rightskip=-3cm
\begin{equation*}
\label{eq:diff}
\hspace*{57pt}
\begin{aligned}
D(\textbf{r},\textbf{v})=\hbar^2 k^2_L\frac{\Gamma}{4}\frac{s_{tot}(\textbf{r},\textbf{v})}{1+s_{tot}(\textbf{r},\textbf{v})}         \quad\quad\quad\quad\text{(A13)}
\end{aligned}
\end{equation*}

\noindent where $\hbar$ is the reduced Planck constant, and 

\begin{equation*}
\label{eq:stot}
\hspace*{75pt}
\begin{aligned}
s_{tot}(\textbf{r},\textbf{v})=\sum_\mathbbm{1}s_\mathbbm{1}(\textbf{r},\textbf{v})             \quad\quad\quad\quad\quad\,\,\text{(A14)}
\end{aligned}
\end{equation*}

\noindent is the total saturation parameter that is a sum of the individual total saturation parameters $s_\mathbbm{1}$, where $\mathbbm{1}=\textsf{3M},\textsf{2M},\textsf{2O}$, or $\textsf{ZS}$ (respectively referring to the 3D MOT, 2D MOT, 2D OM, or ZS). We write $s_\mathbbm{1}$ as the sum of the saturation parameters $s_{q;\mathbbm{1}}$ for the atom's two-level transitions that are driven by, respectively, $\sigma^-$, $\pi$, $\sigma^+$ polarized light:
\begin{equation*}
\label{eq:stot_3DMOT}
\hspace*{65pt}
\begin{aligned}
s_\mathbbm{1}(\textbf{r},\textbf{v})=\sum_{q=\sigma^+,\sigma^-,\pi}s_{q;\mathbbm{1}}(\textbf{r},\textbf{v})      \quad\quad\quad\;\;\text{(A15)}
\end{aligned}
\end{equation*}
These individual parameters are written as sums of parameters for a single beam and atomic transition, expressed by
\begin{equation*}
\label{eq:salpha_q}
\hspace*{45pt}
\begin{aligned}
s_{\alpha',q;\mathbbm{1}}^{\pm,Q}(\textbf{r},\textbf{v})=\frac{p_{\alpha',q;\mathbbm{1}}^{\pm,Q}(\textbf{r}){I}^{\pm}_{\alpha';\mathbbm{1}}(\textbf{r})/I_{sat}}{1+4\frac{(\Delta_\mathbbm{1} -\textbf{k}_{\alpha';\mathbbm{1}}^{\pm}\cdot\textbf{v} - \mu_q(\textbf{r}))^2}{\Gamma^2}}   \quad\;\;\text{(A16)}
\end{aligned}
\end{equation*}
For the 3D MOT, for instance, one has $s_{q;\textsf{3M}}=\sum_{\alpha'=x',y',z'}s^{+,Q}_{\alpha',q;\textsf{3M}}+s^{-,Q}_{\alpha',q;\textsf{3M}}$, where $Q=\sigma^-$ for $\alpha' = \mathsf{x}',\mathsf{y}'$ and $Q=\sigma^+$ for $\alpha'=\mathsf{z}'$.

\vspace*{15pt}
\leftskip=3.6cm
\section*{\hspace{-0.42cm}Appendix B}\label{sec:Appendix_A2}
\leftskip=0.15cm\rightskip=-3cm

This Appendix presents an expression for the mean lifetime of an atom within our $F=0 \rightarrow F'=1$ theoretical model.

From Eq. (A2) in Ref. \cite{2:1} for the steady-state atom number in the excited blue cooling level $5s5p$ $^1P_1$, we can obtain the following expression for the mean atom-lifetime:

\begin{equation*}
\label{eq:tau_simple}
\hspace*{90pt}
\begin{aligned}
\tilde{\tau}=\frac{1}{\Gamma_{23}'}\frac{\Gamma_{34}+\Gamma_{36}}{\Gamma_{34}}\mathcal{E}              \quad\quad\quad\quad\quad\,\,\,\,\text{(B1)}
\end{aligned}
\end{equation*}

\noindent where we include the substitution $\Gamma_{23}\rightarrow\Gamma_{23}'$, where

\begin{equation*}
\label{eq:Gamma_23}
\hspace*{80pt}
\begin{aligned}
\Gamma_{23}'=\frac{\Gamma_{23}}{2}\frac{s_{tot}(\textbf{r},\textbf{v})}{1+s_{tot}(\textbf{r},\textbf{v})}              \quad\quad\quad\quad\quad\text{(B2)}
\end{aligned}
\end{equation*}
is the photon scattering rate for the transition $5s5p\,^1\!P_1\rightarrow5s4d\,^1\!D_2$ due to the blue cycling with the saturation $s_{tot}$ (Eq. \hyperref[eq:stot]{A14}), which we note introduces\;\;dependency\;\;on the

\leftskip=-3cm\rightskip=0.15cm
\begin{figure*}
\vspace*{-240pt}
\hspace*{-73pt}\includegraphics[scale=0.78]{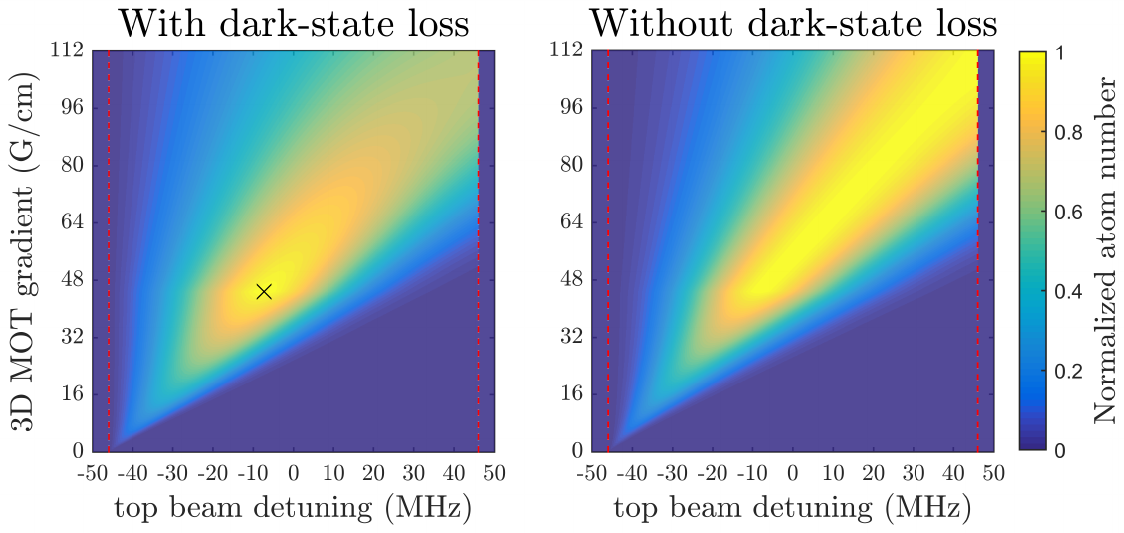}
\vspace*{-223pt}
 \captionsetup{width=1.49\linewidth}
  \caption*{\textbf{Figure 7:} Diagrams showing the normalized 3D MOT atom numbers obtained using our analytical model describing the 3D MOT loading using the moving molasses technique discussed in the main text (Sec. \hyperref[sec:MainResults]{III}). The left diagram uses the full Eq. \hyperref[eq:N]{3} of our model, while the right one excludes the loss to the dark state $5s5p\,^3\!P_0$ ($\Gamma_\text{loss}$ is held at a fixed positive value in Eq. \hyperref[eq:N]{3}). The left dashed line and the cross are as in Fig. \hyperref[fig:4]{4}, while the right dashed line indicates the boundary where the sign of the confinement is flipped. Note that the diagrams span over a larger parameter space than in Fig. \hyperref[fig:4]{4}.}
\label{fig:AppB}
\end{figure*}

\noindent local intensity, magnetic field, and velocity. This substitution is motivated by Eq. (1) in Ref. \cite{2:4}, where one can identify $A_{^1\!P_1\rightarrow^1\!D_2}=\Gamma_{23}=3.9\times10^3$ Hz as the corresponding transition linewidth, and $B_{^1\!D_2\rightarrow^3\!P_2}=\left(\frac{\Gamma_{34}+\Gamma_{36}}{\Gamma_{34}}\right)^{-1}=0.33$ as the branching ratio for the shelving transition $5s4d\,^1\!D_2\rightarrow5s5p\,^3\!P_2$ (with the corresponding linewidth $\Gamma_{34}=6.6\times10^2$ Hz, and $\Gamma_{36}=1.34\times10^3$ Hz being the linewidth for $5s4d\,^1\!D_2\rightarrow5s5p\,^3\!P_1$).

The enhancement factor $\mathcal{E}$ due to the repumping  light acting on the transition $5s5p\,^3\!P_2\rightarrow5p^2\,^3\!P_2$ (cyan 481 nm) has the general expression given by Eq. (A4) in Ref. \cite{2:1}. A recent study \cite{2:2} identified the limit $\mathcal{E}\rightarrow{\sim}27$ (irrespec-

\leftskip=0.15cm\rightskip=-3cm
\noindent tive of the atomic density). We assume this value in our simulations.

We note that as mentioned in Sec. \hyperref[sec:Experiment]{II}, the lifetime of a superparticle in the simulations is determined by multiplying $\tilde{\tau}$ by a number that is predrawn from a unit exponential distribution. 

\vspace*{-10pt}
\leftskip=3.6cm
\section*{\hspace{-0.42cm}Appendix C}\label{sec:Appendix_B}
\leftskip=0.15cm\rightskip=-3cm

Figure \hyperref[fig:AppB]{7} shows two diagrams calculated using our analytical model (Sec. \hyperref[sec:MainResults_i]{III.A}), in order to support our main observations (Sec. \hyperref[sec:MainResults_ii]{III.B}). See the figure text for details.

\end{multicols}


\newpage
\newgeometry{left=1.86in,right=1.86in,top=2.3in,bottom=1.4in}

\title{\vspace{-5.7cm}\textbf{\hspace*{-0.7cm}{\mbox{{\Large Magneto-optical-trap loading in a large optical-access experiment}}}\\ \hspace*{5cm}{\Large (Supplemental Material)}}}
\date{}
\maketitle

\begin{multicols}{2}\setlength{\columnsep}{2pt}

\leftskip=-3cm\rightskip=0.15cm
We here demonstrate our simulation tool's ability to improve the performance of an effusive oven experiment. We discuss how the 2D MOT loading can be enhanced by introducing changes to the 2D OM and the ZS. Also, we numerically verify the optimal angles for the radial beams in the 3D MOT. 

Two important ways to manipulate the 2D OM performance include optimizing the beam power and the beam size (aperture and waist) [37]. The former is, however, resource intensive, considering the gains diminish rapidly due to the atom saturation while other parts of the setup (e.g., the 2D and 3D MOTs) may benefit more from the available laser power. In Fig. \hyperref[fig:S.1]{S.1}, we study the effects of increased beam power for different waist sizes at fixed aperture sizes. The data points connected with a straight line are obtained for the original aperture and beam waist (Tab. 1 case 4), while the data points connected with other kinds of lines are for a $4$ times larger aperture (requiring two-inch optics) and different beam waists (dashed for $1$ times the original, dash-dotted for $2$, and dotted for $3$). We observe first that enlarging the aperture can result in an increased 2D MOT loading rate, depending on both the beam waist and the 2D OM power. An increase in this rate is a consequence of the atoms being exposed longer to the transverse cooling, provided the intensity is sizeable throughout the cooling region with the chosen beam waist.  Indeed, as observed for lower intensities (below 30 mW), extending the waist can either be advantageous (dashed and dash-dotted lines) due to effectively improved cooling, or detrimental (dotted line) due to effectively diminished cooling. For larger intensities, on the other hand, the saturation effect is present, such that wider waists result in the extension of a region of efficient cooling. In this case, it is particularly important to choose a waist that is not too narrow, so this effect can be taken full\;\,advantage\;\;of\;\,(observe the\;\;dotted\;\;line 

\leftskip=0.15cm\rightskip=-3cm
\noindent crossing the dashed one above 30 mW). Overall, the observations here agree with intuitive expectations, further validating the numerical model.

\vspace*{-155pt}
\hspace*{-80pt}\includegraphics[scale=0.64]{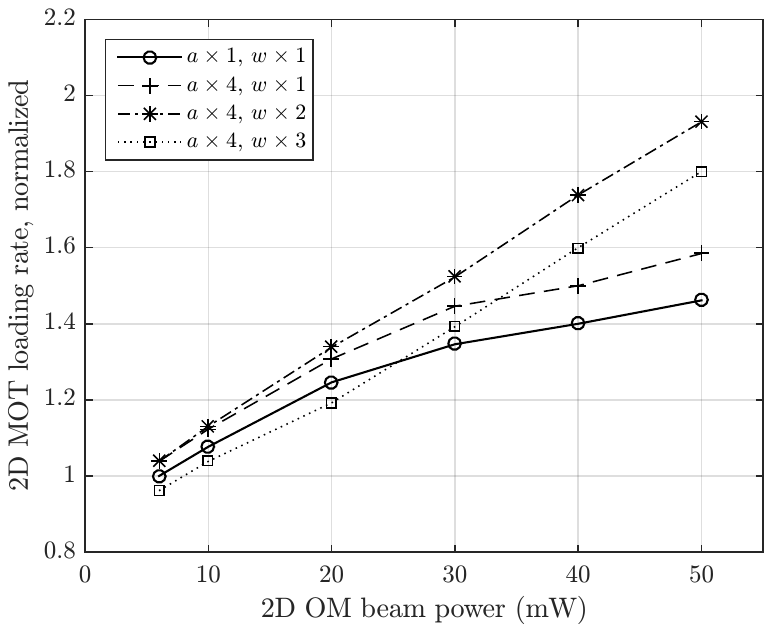}\label{fig:S.1}
\vspace*{-157pt}
\begin{footnotesize}
\begin{spacing}{1}
\vspace*{-3pt}\noindent{\textbf{Figure S.1:} Numerical data showing the 2D MOT loading rate versus the 2D OM beam power (individual). Different aperture radii $a$ and beam waist radii $w$ of the 2D OM are considered. The aperature size "$a\times1$" and the waist size "$w\times1$" are from the original simulations, where $6$ mW of 2D OM beam power is also used (Tab. 1 case 4). The loading normalization is done with respect to this case.\newline}
\end{spacing}
\end{footnotesize}

Our ZS modeling (Eq. A8) has been derived with a purpose of tailoring a magnetic field to a given experimental environment, and we outline here how this can be achieved. With the initial oven conditions, ZS beam and 2D MOT parameters specified, one employs an algorithm that iteratively adjusts the slopes of the linear segments  composing the field profile, with the aim of increasing the probability density of the atom speeds within the capture-speed range of the 2D MOT. To remove spurious fluctuations, each iteration step uses the same initial
\end{multicols}

\newgeometry{left=1.86in,right=1.86in,top=0.85in,bottom=0.8in}
\begin{multicols}{2}\setlength{\columnsep}{2pt}
\leftskip=-3cm\rightskip=0.15cm
\noindent atom position and speed distributions with the diffusion (stemming from the radiation pressure; App. A1.B) turned off. In Fig. \hyperlink{page.2}{S.2}, we display the initial speed profile (light gray), and the final speed profile at the 2D MOT location (dark gray), which can be achieved in the iteration process. A concentrated probability density is seen to occur close to 0 m/s for the final profile (see the inset), corresponding to the atoms that have been captured by the 2D MOT. The employed field profile stems from our experiment where it more than quadruples the 2D MOT loading rate (see Tab. I), but it has not been optimized using the numerical approach outlined above. We expect it to yield accurate optimization results, given the closely matching numerical and experimental 2D MOT loading rates obtained with help of the segmentation, and possibly achieve a 20-fold enhancement with appropriate ZS stage modifications [37, 40]. We note that our numerical approach is similar to the one extensively tested in Ref. [40], with the main difference there being that the motion is in 1D and the radial expansion of the oven flux is approximated to yield an effective 3D simulation. We note that although the lower (actual) dimensionality will result in greater computational efficiency, this difference may lead to decreased accuracy compared to our approach.

Finally, we verify for the 3D MOT side beams their optimal angles with respect to the radial axis perpendicular to the objectives axis. In Fig. \hyperlink{page.2}{S.3}, we display the numerical results for the 3D MOT atom numbers versus the beams angle (mirrored). As can be seen, the $25\,^{\circ}$ angle yields the highest number given our experimental constrains (dashed line limit). While larger angles can result in even higher numbers (see right of the dashed line), the beams may be clipped by our objectives [refer to the inset in Fig. 2(a)]. The optimum angle is found to be greater, at $30\,^{\circ}$, which can be explained to be a combination of a large repumping light volume in the science region (thus bringing many shelved atoms back into the cooling cycle) and the radial trapping forces being closer to perpendicular (thus resulting in more balanced forces). The resulting increase in the atom number is, however, marginal (by ${\sim}25\,\%$), and it is relatively robust against changes within a wide range of angles (from $20\,^{\circ}$ to $45\,^{\circ}$). Note that the findings here are not universal due to their dependence on the cell geometry. Implicitly, however, we are made aware that the optimum loading is in general not\;\;achieved\;\;at\;\;the\;\;standard $90\,^{\circ}$ angle\;\,between\;\,beam 

\leftskip=0.15cm\rightskip=-3cm
\noindent axes (corresponding to $45\,^{\circ}$ in Fig. \hyperlink{page.2}{S.3}). 

\vspace*{-163pt}
\hspace*{-80pt}\includegraphics[scale=0.64]{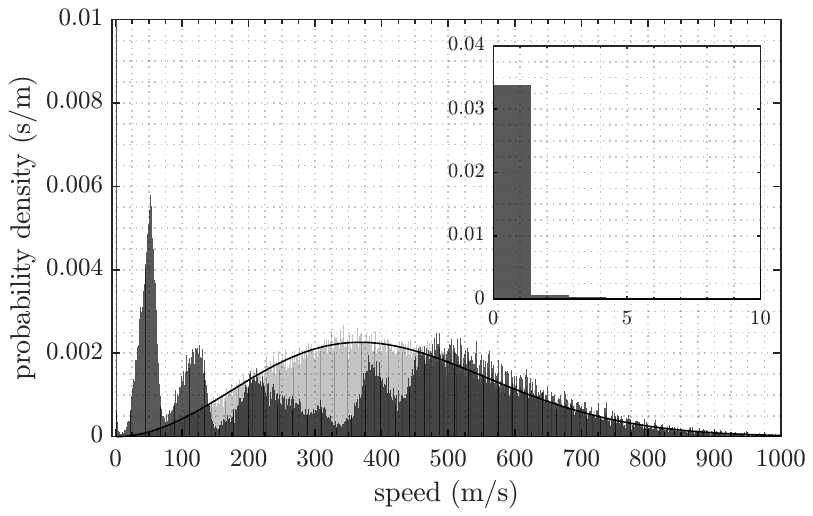}\label{fig:S.2}
\vspace*{-178pt}
\begin{footnotesize}
\begin{spacing}{1}
\vspace*{-3pt}\noindent{\textbf{Figure S.2:} Numerical data showing atom speed histograms before the Zeeman slowing (light gray) and at the 2D MOT location after the slowing (dark gray). The solid line displays the Maxwell-Boltzmann distribution for $440\,^{\circ}$C (as used in the original simulations; see Sec. II.A). The inset zooms in on the capturable speeds of the 2D MOT, showing a concentrated probability density near 0 m/s due to the trapping.\newline}
\end{spacing}
\end{footnotesize}

\vspace*{-158pt}
\hspace*{-80pt}\includegraphics[scale=0.64]{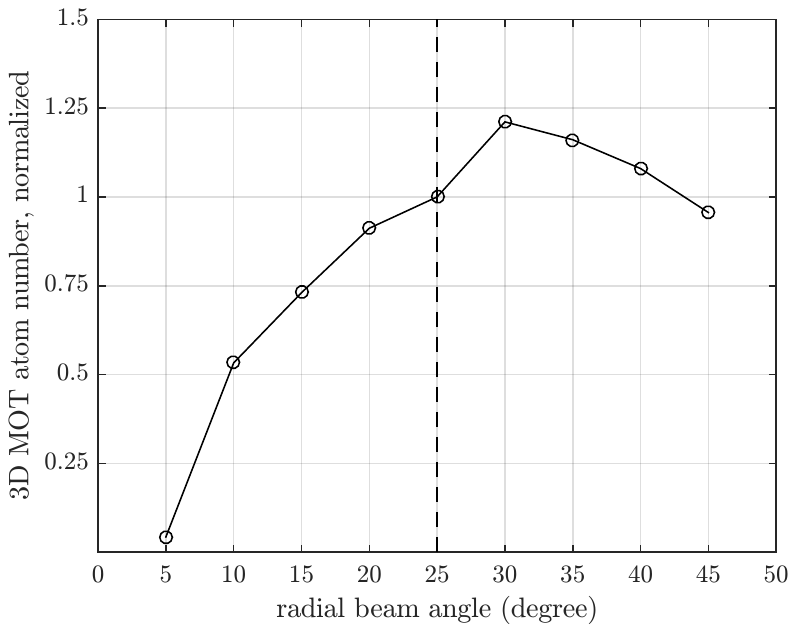}\label{fig:S.3}
\vspace*{-155pt}
\begin{footnotesize}
\begin{spacing}{1}
\vspace*{-3pt}\noindent{\textbf{Figure S.3:} Numerical data showing the 3D MOT atom number versus the angle of the side beams mirrored with respect to the radial axis perpendicular to the objectives axis [refer to the inset in Fig. 2(a)]. The vertical dashed line indicates the limit beyond which our objectives may clip the beams. The normalization is done with respect to the point at the $25\,^{\circ}$ angle, as used in the experiment (see Sec. II.D). The remaining simulation parameters are the same as in Fig. 4, where the greatest atom number is obtained.\newline}
\end{spacing}
\end{footnotesize}

\end{multicols}


\end{document}